\documentclass[nofootinbib, reprint, preprintnumbers, amsmath, amssymb, amsfonts, aps, pra, superscriptaddress, floatfix]{revtex4-2}
\usepackage{mathtools}
\usepackage{url}
\usepackage{booktabs}
\usepackage{graphicx}
\usepackage{dcolumn}
\usepackage{bm}
\usepackage{physics}
\usepackage{placeins}
\usepackage{color}
\usepackage{braket}
\usepackage[normalem]{ulem}
 \usepackage{enumerate}
\usepackage{enumitem}
\usepackage[colorlinks=true,linkcolor=blue,citecolor=blue,urlcolor=blue]{hyperref}
\usepackage{mathrsfs}
\usepackage{bbm}
\usepackage{siunitx}
\usepackage{comment}

\begin{document}

\title{
Experimental Measurement and Theoretical Analysis of Energy Relaxation Rates of an Interacting Two-Qubit System on a D-Wave Quantum Annealer
}
\author{Yusei Amano}
\affiliation{Tokyo University of Science, 1-3 Kagurazaka, Shinjuku-ku, Tokyo, Japan}

\author{Sorato Suzuki}
\affiliation{Tokyo University of Science, 1-3 Kagurazaka, Shinjuku-ku, Tokyo, Japan}

\author{Hiroki Kuji}
\affiliation{Tokyo University of Science, 1-3 Kagurazaka, Shinjuku-ku, Tokyo, Japan}

\author{Tetsuro Nikuni}
\affiliation{Tokyo University of Science, 1-3 Kagurazaka, Shinjuku-ku, Tokyo, Japan}

\author{Takashi Imoto}
\affiliation{RIKEN, 2-1 Hirosawa, Wako, Saitama 351-0198, Japan}

\author{Yuichiro Matsuzaki}
\email{ymatsuzaki872@g.chuo-u.ac.jp}
\affiliation{Department of Electrical, Electronic, and Communication Engineering, Faculty of Science and Engineering, Chuo University, 1-13-27 Kasuga, Bunkyo-ku, Tokyo, Japan}

\date{\today}

\begin{abstract}
In D-Wave quantum annealers, various properties of the ground state have been clarified, whereas
the correspondence between theory and experiment for energy relaxation rates in multiqubit excited states remains insufficiently understood. Here, we measured the energy relaxation rates of
the first excited states in single-qubit systems and interacting two-qubit systems using a D-Wave
quantum annealer. We analyzed the measured relaxation rates using a Gorini-Kossakowski-Sudarshan-Lindblad master
equation with independent local $\sigma^z$-type noise channels. The calculated relaxation rates reproduced
the overall trends observed experimentally, supporting the model at a qualitative level. We then
used the relaxation rates measured for the uncoupled single-qubit systems to calibrate the local
relaxation parameters and predict the relaxation rates of the interacting two-qubit systems. The
predicted rates remained within a factor of approximately four of the measured rates. These results
show that the relaxation measurements on individual qubits can provide a practical estimate of
relaxation in small interacting quantum systems and may help clarify relaxation mechanisms in
programmable quantum annealers.
\end{abstract}

\maketitle

\section{Introduction}
Quantum annealing has been developed as a useful method
for solving combinatorial optimization problems \cite{Finnila1994,Kadowaki1998,Farhi2000,Farhi2001,
Santoro2002,Das2008,AlbashLidar2018}.
D-Wave Quantum Inc. has developed quantum annealers in which thousands of
rf-SQUID flux qubits are integrated
\cite{Harris2010UnitCell,Johnson2011,King2023Nature}.
A broad class of combinatorial optimization problems can be mapped onto the problem of
finding the ground state of an Ising Hamiltonian
\cite{Barahona1982,Lucas2014}.

In addition, programmable quantum annealers can be used not only as tools for
combinatorial optimization but also as quantum simulators for investigating
dynamics, phase transitions, and topological phenomena in quantum many-body
systems under equilibrium and nonequilibrium conditions
\cite{King2022,King2018Nature,Kairys2020PRXQuantum,
Harris2018Science,Zhou2021PRB,Lanting2014,Albash2015}.
In particular, entanglement has been demonstrated
in selected few-qubit systems implemented on a D-Wave
quantum processor
\cite{Lanting2014,Albash2015}.
King et
al. also demonstrated coherent many-body dynamics in
a programmable Ising chain, showing that D-Wave quantum annealers can be operated as quantum simulators
of interacting quantum systems \cite{King2022}.

For such devices to be used as quantitative quantum
simulators, it is necessary to understand the influence of
environmental noise and dissipation in addition to coherent Hamiltonian dynamics. To date, the noise and coherence properties of superconducting flux qubits, including
rf-SQUID flux qubits, have been investigated from both
theoretical and experimental perspectives \cite{Yoshihara2006,Novikov2018,Kakuyanagi2007,Harris2008,Amin2008PRL,Harris2010Robust}. For
example, Harris et al. characterized low-frequency flux
noise in rf-SQUID flux qubits \cite{Harris2008}. These studies motivate the use of local $\sigma^{z}_k$-type noise channels for flux qubits. However, the energy-relaxation rate between specific eigenstates also depends on the noise spectrum at the
transition frequency and on the corresponding transition
matrix elements \cite{Breuer2002}.

Excited-state relaxation on D-Wave quantum annealers has also been investigated \cite{Imoto2023}.
Imoto \textit{et al.} demonstrated a reverse-annealing method for measuring the
energy-relaxation times of first excited states and
compared a single-qubit system with a fully connected
four-qubit transverse-field Ising system. They observed
a substantially longer relaxation time in the four-qubit
system and attributed this behavior to suppressed local transition matrix elements. Building on
this work, the question addressed in the present study
is whether relaxation measurements of the constituent
physical qubits, performed with the interaction turned
oﬀ, can be used as reference data to predict the relaxation rate of the interacting system.

In this study, we test such a calibration-based prediction for interacting two-qubit systems on a D-Wave
quantum annealer. Specifically, we measure the relaxation rates from the first excited state to the ground
state for the same physical qubits operated individually
and as interacting pairs. We analyze the measured relaxation rates within a Gorini–Kossakowski–Sudarshan–Lindblad (GKSL) master equation with independent local $\sigma^{z}_k$-type noise channels~\cite{Gorini1976,Lindblad1976}. First, we examine
the relationship between the measured single-qubit relaxation rates and the corresponding squared transition
matrix elements. We then use the single-qubit measurements to infer the local transition-rate coeﬃcients and predict the relaxation rates of the interacting two-qubit
systems. The single-qubit rates approximately follow
the transition-matrix-element scaling expected from the
model. For the interacting systems, the predictions reproduce the overall trend and remain within a factor of
approximately four of the measured rates over the investigated parameter range, although the model systematically underestimates the rates in the small-transition-matrix-element region.

The remainder of this paper is organized as follows. In Sec.~II, we introduce the theoretical model based on the GKSL master equation and describe the method used to predict the relaxation rates of the interacting two-qubit systems from the single-qubit measurements. In Sec.~III, we describe the reverse-annealing protocol and the experimental procedure used to determine the energy relaxation rates. In Sec.~IV, we present the experimental results and compare the predicted and measured relaxation rates. Finally, Sec.~V summarizes the main conclusions of this study.

\section{Theoretical Model and Prediction Method}
In this section, we formulate the energy relaxation rate
based on a GKSL equation and describe a method for
predicting the relaxation rate of an interacting two-qubit
system from the experimentally measured single-qubit
relaxation rates.
\subsection{Energy Relaxation Rate Based on the GKSL Master Equation}
In this study, we use a GKSL master equation with independent local noise channels to evaluate the energy relaxation rates of excited states. As a noise model, we assume independent $\sigma_k^z$-type noise acting on each qubit. Here, $\sigma_k^z$ ($\sigma_k^x$) denotes the Pauli-$z$(x) operator acting on the $k$th qubit. This assumption is motivated by the fact that flux noise is one of the dominant noise sources in superconducting flux qubits \cite{Yoshihara2006,Kakuyanagi2007,Harris2008}. 
These studies primarily characterize low-frequency flux noise; they motivate modeling the system–bath coupling as local longitudinal noise acting through the operators $\sigma_k^z$, but do not determine the frequency-dependent transition-rate coeﬃcients near the transition frequency of approximately 1.5 GHz studied here. We neglect noise correlations between diﬀerent qubits.
We further employ Born-Markov approximations and rotating wave approximations for the decoherence process.
Under these assumptions, the time evolution of the density matrix $\rho(t)$ is described by the following equation.
\begin{align}
\frac{d\rho(t)}{dt}
&= -i[H,\rho(t)]
\nonumber\\
&\quad
+\sum_k\sum_\omega \gamma_k(\omega)
\Bigl[
L_k(\omega)\rho(t)L_k^\dagger(\omega)
\nonumber\\
&\qquad
-\frac{1}{2}
\left\{
L_k^\dagger(\omega)L_k(\omega),
\rho(t)
\right\}
\Bigr].
\label{eq:gksl_master_new}
\end{align}
Here, $H$ is the system Hamiltonian of the qubit system under consideration. The quantity $\gamma_k(\omega)$ is the frequency-dependent transition-rate coeﬃcient for the $k$th qubit; it incorporates the
relevant noise spectrum and system–bath coupling
strength. Throughout this paper, we set $\hbar=1$ in the theoretical expressions.
Energies quoted in GHz are expressed in frequency units
as $E/(2\pi\hbar)$. The noise operator $L_k(\omega)$ represents the component associated with transitions whose energy difference is $\omega$. When the eigenvalues and eigenstates of $H$ are written as $H|\psi_\epsilon\rangle=\epsilon|\psi_\epsilon\rangle$, $L_k(\omega)$ is defined as follows \cite{Breuer2002}:
\begin{equation}
L_k(\omega)
=
\sum_{\epsilon'-\epsilon=\omega}
|\psi_\epsilon\rangle
\langle\psi_\epsilon|
\sigma_k^z
|\psi_{\epsilon'}\rangle
\langle\psi_{\epsilon'}|.
\label{eq:Akomega}
\end{equation}
We further assume the low-temperature limit and retain only the downward transition from the first excited state $|\Phi^{}_{1\mathrm{st}}\rangle$ to the ground state $|\Phi^{}_{\mathrm{gs}}\rangle$. Denoting the energy difference between the first excited state and the ground state of the system under consideration by 
$\Delta E$, 
we retain only the term with $\omega = \Delta E$ in $L_k(\omega)$.
For the transition from the
first excited state to the ground state, the corresponding jump operator for the $k$th qubit is
\begin{align}
L_k(\Delta E)
&= |\Phi_{\mathrm{gs}}\rangle \langle\Phi_{\mathrm{gs}}|
\sigma^{z}_k|\Phi_{1\mathrm{st}}\rangle
\langle\Phi_{1\mathrm{st}}| \notag \\
&= \langle\Phi_{\mathrm{gs}}|\sigma^{z}_k|\Phi_{1\mathrm{st}}\rangle
|\Phi_{\mathrm{gs}}\rangle \langle\Phi_{1\mathrm{st}}|.
\label{eq:AkDeltaE}
\end{align}
In the above expression, the noise operator $|\Phi^{}_{\mathrm{gs}}\rangle\langle\Phi^{}_{1\mathrm{st}}|$ represents the transition from the first excited state to the ground state, while the transition matrix element $\langle\Phi^{}_{\mathrm{gs}}|\sigma^{z}_k|\Phi^{}_{1\mathrm{st}}\rangle$ characterizes the strength of this transition. Substituting this expression into the GKSL equation and taking the matrix element with
respect to $|\Phi^{}_{\mathrm{1st}}\rangle$, we obtain
\begin{equation}
\frac{dP_{\mathrm{1st}}(t)}{dt}
=
-\Gamma P_{\mathrm{1st}}(t),
\label{eq:population_decay}
\end{equation}
where
\begin{equation}
P_{\mathrm{1st}}(t)
=
\langle\Phi_{\mathrm{1st}}|
\rho(t)
|\Phi_{\mathrm{1st}}\rangle.
\label{eq:first_excited_population}
\end{equation}
The energy relaxation rate from the first excited state to the ground state is given by
\begin{equation}
\Gamma^{} = \sum_{k=1}^{N} \gamma_k(\Delta E) \left| \langle\Phi^{}_{\mathrm{gs}}| \sigma_k^{z} |\Phi^{}_{1\mathrm{st}}\rangle \right|^2 .
\label{gensui}
\end{equation}
Here, $N$ denotes the number of qubits in the system.
\subsection{Prediction of the Interacting Two-Qubit Relaxation Rate Using Independent Single-Qubit Reference Measurements}

In this subsection, we consider a system consisting of two qubits. We refer to the system with the qubit--qubit interaction turned off as the non-interacting two-qubit system, and the system with the interaction turned on as the interacting two-qubit system. The experimentally measured relaxation rates of the individual qubits in the non-interacting two-qubit system are used to predict the relaxation rate of the interacting two-qubit system.
\subsubsection{Independent Single-Qubit Reference Measurements}

\begin{figure}[htbp]
    \centering
    \includegraphics[width=0.85\linewidth]{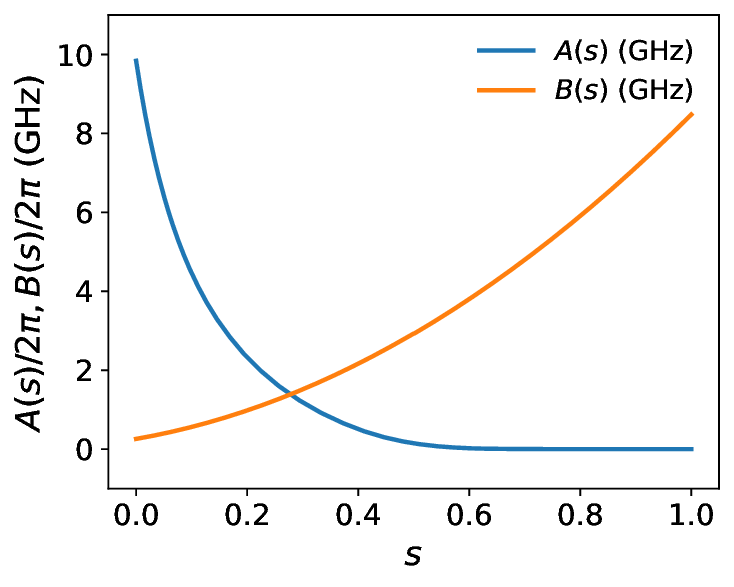}
    \caption{
    Annealing schedule functions $A(s)$ and $B(s)$ for the
    D-Wave Advantage 4.1 quantum annealer.
    The curves were generated by the authors using the
    device-specific annealing-schedule data provided by
    D-Wave Quantum Inc.~\cite{DWaveSchedule}.
    Here, $s$ is the normalized annealing parameter ranging
    from 0 to 1. The function $A(s)$ represents the transverse-field
    energy scale, whereas $B(s)$ sets the
energy scale multiplying the problem Hamiltonian, including
the longitudinal-field and Ising-interaction terms.
    }
    \label{asbs.fig}
\end{figure}

First, we consider the non-interacting system, in which the interaction between the qubits is turned off. In this case, the non-interacting two-qubit Hamiltonian is given by
\begin{equation}
H^{(\mathrm{off})}(s)
=
-\frac{A(s)}{2}
\left(\sigma_1^x+\sigma_2^x\right)
+\frac{B(s)h}{2}
\left(\sigma_1^z+\sigma_2^z\right).
\label{eq:H_off}
\end{equation}
Here, $A(s)$ and $B(s)$ are device-dependent schedule coefficients that control the relative strengths of the transverse-field and 
longitudinal field,
as shown in Fig.~\ref{asbs.fig}. The parameter $h$ denotes the strength of the longitudinal field applied to each qubit, and $s$ is  the dimensionless annealing parameter. 

For notational convenience, let $|g\rangle$ and $|e\rangle$
denote the ground and excited states, respectively,
of each single-qubit Hamiltonian at the hold point,
and write the corresponding two-qubit states as $|ee\rangle,\quad |eg\rangle,\quad |ge\rangle,\quad|gg\rangle$.
Here, $|eg\rangle$ ($|ge\rangle$) denotes the state in which qubit 1 is in the excited (ground) state and qubit 2 is in the ground (excited) state. Similarly, $|ee\rangle$ ($|gg\rangle$) denotes the state in which both qubits are in the excited (ground) state.

We evaluate the two relaxation processes in the non-interacting system,
$|eg\rangle \rightarrow |gg\rangle$ and
$|ge\rangle \rightarrow |gg\rangle$,
separately.
The experimentally measured energy relaxation rates corresponding to these
processes are denoted by
$\Gamma^{(\mathrm{off})}_{1,\mathrm{exp}}$
and
$\Gamma^{(\mathrm{off})}_{2,\mathrm{exp}}$,
respectively.
Here,
$\Gamma^{(\mathrm{off})}_{1,\mathrm{exp}}$
corresponds to the relaxation process
$|eg\rangle \rightarrow |gg\rangle$,
whereas
$\Gamma^{(\mathrm{off})}_{2,\mathrm{exp}}$
corresponds to
$|ge\rangle \rightarrow |gg\rangle$.
The corresponding squared transition matrix elements are defined as 
\begin{align}
M^{(\mathrm{off})}_1
&=
\left|
\langle gg|\sigma_1^z|eg\rangle
\right|^2
\label{eq:M_off_1},
\\
M^{(\mathrm{off})}_2
&=
\left|
\langle gg|\sigma_2^z|ge\rangle
\right|^2.
\label{eq:M_off_2}
\end{align}

In this study, we assume that the eﬀective transition-rate coeﬃcients remain approximately unchanged during the measurements and can be transferred between the non-interacting and interacting systems when the transition frequencies are approximately the same. Under the assumptions of
the present model, the experimentally measured energy
relaxation rates of qubits 1 and 2 are given by the
following expressions:
\begin{align}
\Gamma^{(\mathrm{off})}_{1,\mathrm{exp}}
&=
\gamma_1(\Delta E) M^{(\mathrm{off})}_1,
\quad
\Gamma^{(\mathrm{off})}_{2,\mathrm{exp}}
=
\gamma_2(\Delta E) M^{(\mathrm{off})}_2.
\label{eq:Gamma_off_2}
\end{align}
Thus, the eﬀective transition-rate coeﬃcient for each qubit can be inferred from the experimentally measured relaxation rates in the non-interacting system as follows:
\begin{align}
\gamma_1(\Delta E)
=
\frac{\Gamma^{(\mathrm{off})}_{1,\mathrm{exp}}}
{M^{(\mathrm{off})}_1},
\qquad
\gamma_2(\Delta E)
=
\frac{\Gamma^{(\mathrm{off})}_{2,\mathrm{exp}}}
{M^{(\mathrm{off})}_2}.
\label{eq:gamma_2}
\end{align}
\subsubsection{Prediction of the Two-Qubit Relaxation Rate in the Interacting System}

We next consider the system that includes interaction between the two qubits.
The corresponding two-qubit Hamiltonian is given by
\begin{align}
H^{(\mathrm{on})}(s)
&=
-\frac{A(s)}{2}
\left(\sigma_1^x+\sigma_2^x\right)
\nonumber\\
&\quad
+\frac{B(s)}{2}
\left(
h\sigma_1^z+h\sigma_2^z
+J_{12}\sigma_1^z\sigma_2^z
\right).
\label{eq:H_on}
\end{align}

In this study, we employ a ferromagnetic interaction with coupling strength $J_{12}=-1$.
For the ideal interacting two-qubit Hamiltonian considered here,
the transverse field mixes different Ising configurations, and can generate entanglement in the
low-energy eigenstates.
The same
state mixing also generates nonzero local $\sigma^z_k$ transition
matrix elements between the ground and first excited
states. In Appendix~\ref{app:concurrence}, we analyze these eﬀects in
the weak-transverse-field regime and show that, for
fixed longitudinal field, both the concurrence and the
local transition matrix element arise at second order
in the transverse-field strength. We further compare
the perturbative results with exact diagonalization and
discuss the correlated variation of these quantities
along the parameter path considered in this study.
The concurrence considered here characterizes the pure
eigenstates of the ideal Hamiltonian and should not be
interpreted as a direct measurement of entanglement in
the actual open-system state. A related perturbative
analysis of the suppression of local transition matrix
elements in a four-qubit system was presented by Imoto \textit{et al.} \cite{Imoto2023}.

We denote the ground state and the first excited state of the interacting system by
$\ket{\Phi_{\mathrm{gs}}^{(\mathrm{on})}}$
and
$\ket{\Phi_{1\mathrm{st}}^{(\mathrm{on})}}$,
respectively.
The squared transition matrix elements between the first excited state and the ground state are defined as
\begin{align}
M^{(\mathrm{on})}_1
&=
\left|
\bra{\Phi_{\mathrm{gs}}^{(\mathrm{on})}}
\sigma_1^z
\ket{\Phi_{1\mathrm{st}}^{(\mathrm{on})}}
\right|^2,
\label{eq:M_on_1}
\\
M^{(\mathrm{on})}_2
&=
\left|
\bra{\Phi_{\mathrm{gs}}^{(\mathrm{on})}}
\sigma_2^z
\ket{\Phi_{1\mathrm{st}}^{(\mathrm{on})}}
\right|^2.
\label{eq:M_on_2}
\end{align}
Because the interacting two-qubit Hamiltonian considered here is symmetric under the exchange of qubits 1 and 2, these squared transition matrix elements are identical:
\begin{equation}
M^{(\mathrm{on})}_1
=
M^{(\mathrm{on})}_2.
\label{eq:M_on_symmetry}
\end{equation}
Therefore, only one value of $M_k^{(\mathrm{on})}$ is listed for each measurement condition in Table~\ref{tab:measurement_params}(b).

Under the assumptions of the present model, the predicted energy relaxation rate of the interacting two-qubit system is given by
\begin{equation}
\Gamma^{(\mathrm{on})}_{\mathrm{pred}}
=
\gamma_1(\Delta E)M^{(\mathrm{on})}_1
+
\gamma_2(\Delta E)M^{(\mathrm{on})}_2.
\label{eq:Gamma_on_pred_gamma}
\end{equation}
Since the effective transition-rate coefficients $\gamma_k(\Delta E)$ depend on the energy difference $\Delta E$, we adjust $\Delta E$ to be approximately the same in the non-interacting and interacting measurements. We therefore assume that the effective transition-rate coefficients $\gamma_k(\Delta E)$ inferred from the single-qubit measurements using Eq.~(\ref{eq:gamma_2}) can be transferred to the same physical qubits in the interacting system.

Under this assumption, substituting Eq.~(\ref{eq:gamma_2}) into Eq.~(\ref{eq:Gamma_on_pred_gamma}), we predict the energy relaxation rate of the interacting two-qubit system as
\begin{align}
\Gamma^{(\mathrm{on})}_{\mathrm{pred}}
&=
\frac{\Gamma^{(\mathrm{off})}_{1,\mathrm{exp}}}
{M^{(\mathrm{off})}_1}
M^{(\mathrm{on})}_1
+
\frac{\Gamma^{(\mathrm{off})}_{2,\mathrm{exp}}}
{M^{(\mathrm{off})}_2}
M^{(\mathrm{on})}_2 .
\label{eq:Gamma_on_pred_general}
\end{align}
In this study, we compare the predicted energy relaxation rate of the interacting two-qubit system,
$\Gamma^{(\mathrm{on})}_{\mathrm{pred}}$,
obtained from Eq.~(\ref{eq:Gamma_on_pred_general}),
with the experimentally measured energy relaxation rate in the interacting system,
$\Gamma^{(\mathrm{on})}_{\mathrm{exp}}$.

\section{Experimental Method}

In this section, we describe the experimental procedure used to
measure the relaxation from the first excited state during quantum
annealing using the D-Wave Advantage 4.1 system.

Following the method of Imoto \textit{et al.}~\cite{Imoto2023},
we employed reverse annealing with the first excited state of the
system Hamiltonian as the initial state.
We performed measurements for both single-qubit systems and
interacting two-qubit systems.
For the interacting two-qubit systems, we used a ferromagnetic
coupling with $J_{12}=-1$.

For each measurement condition, the longitudinal field $h$ and the
hold point $s_d$ were chosen such that the energy gap between the
ground state and the first excited state was approximately
$\Delta E=1.5~\mathrm{GHz}$.
The detailed values of $h$, $s_d$, $\Delta E$, the squared transition
matrix elements, and the number of shots are summarized in
Table~\ref{tab:measurement_params}.

For each parameter set listed in Table~\ref{tab:measurement_params},
the Hamiltonian at the hold point $s=s_d$ was numerically diagonalized
to calculate the energy gap $\Delta E$ and the corresponding squared
transition matrix elements.
The parameters $h$ and $s_d$ were selected such that
$\Delta E$ was approximately $1.5~\mathrm{GHz}$ in both the
non-interacting and interacting systems.
The calculated $M_k^{(\mathrm{off})}$ values were used together with
the experimentally measured single-qubit relaxation rates
$\Gamma_{k,\mathrm{exp}}^{(\mathrm{off})}$ to determine the effective
single-qubit relaxation contributions, whereas
$M_k^{(\mathrm{on})}$ was used to predict the relaxation rate of the
interacting two-qubit system using
Eq.~(\ref{eq:Gamma_on_pred_general}).
The number of shots listed in Table~\ref{tab:measurement_params}
was used to evaluate the survival probability and its statistical
uncertainty.

\begin{figure}[htbp]
    \centering
    \includegraphics[width=0.85\linewidth]{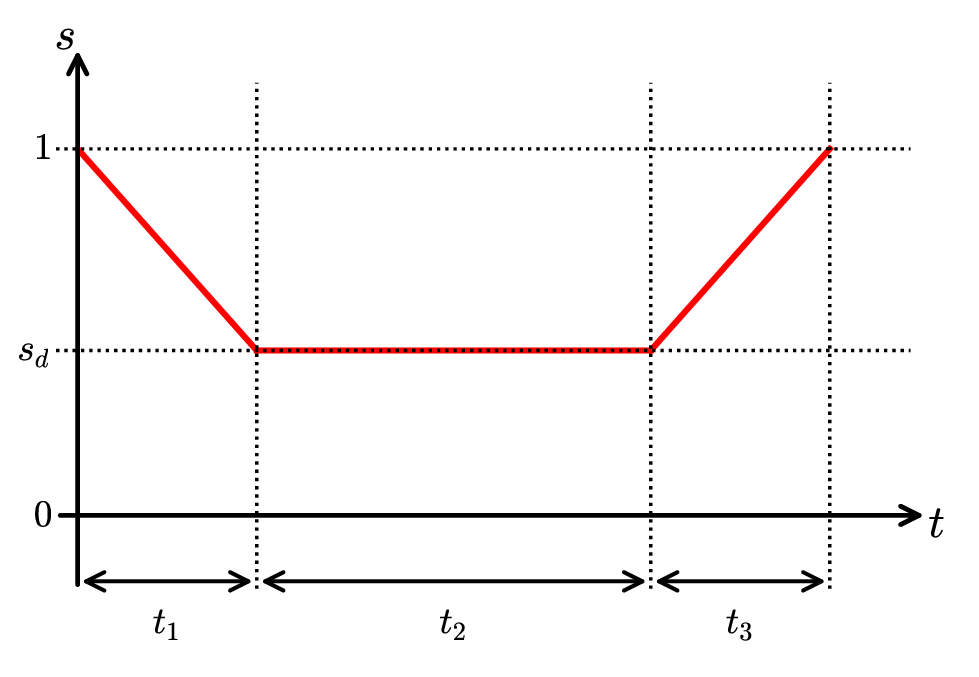}
    \caption{
    Reverse-annealing schedule used to measure the relaxation from the
    first excited state.
    The annealing parameter $s$ is linearly decreased from $s=1$ to the
    hold point $s=s_d$ over a time $t_1$, kept at $s=s_d$ for a hold
    time $t_2$, and then linearly increased back to $s=1$ over a time
    $t_3$.
    }
    \label{schedule.fig}
\end{figure}

The reverse-annealing schedule used in the experiment is shown in
Fig.~\ref{schedule.fig}.
First, the spin configuration corresponding to the first excited state
is prepared at $s=1$.
The annealing parameter $s$ is then linearly decreased from $s=1$ to
the hold point $s=s_d$ over a time $t_1$.
Next, the system is held at $s=s_d$ for a hold time $t_2$, during which
relaxation from the first excited state to the ground state occurs.
After the hold time $t_2$, $s$ is linearly increased from $s_d$ back to
$s=1$ over a time $t_3$.
Finally, the spin configuration is measured in the computational
($\sigma^z$) basis.

For each measurement condition, the above protocol was repeated for
various values of the hold time $t_2$ while the other parameters were
kept fixed.
In this study, we set
$t_1=t_3=1~\mu\mathrm{s}$, for which nonadiabatic transitions are
sufficiently suppressed under the experimental conditions considered
here.
The survival probabilities obtained by varying $t_2$ and the procedure
used to extract the relaxation rates are presented in Sec.~IV.

\section{Results}
The main objective of this section is to examine whether the energy relaxation rate of the interacting two-qubit system can be predicted from the relaxation measurements of the constituent single qubits using Eq.~(\ref{eq:Gamma_on_pred_general}). First, we describe how the experimental relaxation rates are extracted from the measured survival probabilities. We then examine the dependence of the single-qubit and two-qubit relaxation rates on the corresponding squared transition matrix elements. Finally, we use the single-qubit measurements as reference data to predict the two-qubit relaxation rates and compare the predicted values with the experimentally measured ones.
In this study, we analyzed three pairs of physically coupled qubits on the D-Wave Advantage 4.1 system.
The qubits 1 and 2 introduced in the previous section are assigned to the two physical qubits constituting each pair.
Specifically, pair A is defined as $(A_1,A_2)=(2548,5338)$, pair B as $(B_1,B_2)=(4016,2240)$, and pair C as $(C_1,C_2)=(4615,1959)$.
Here, the numbers in parentheses denote the identification labels of the physical qubits used in the single-qubit measurements.

In the following, the pair index and the measurement-condition index are omitted for simplicity unless explicitly required. The fitting parameters are determined independently for each physical qubit or qubit pair, as specified below.

\subsection{Energy-relaxation curves}
\begin{figure*}[t]
\centering
\includegraphics[
width=\textwidth
]{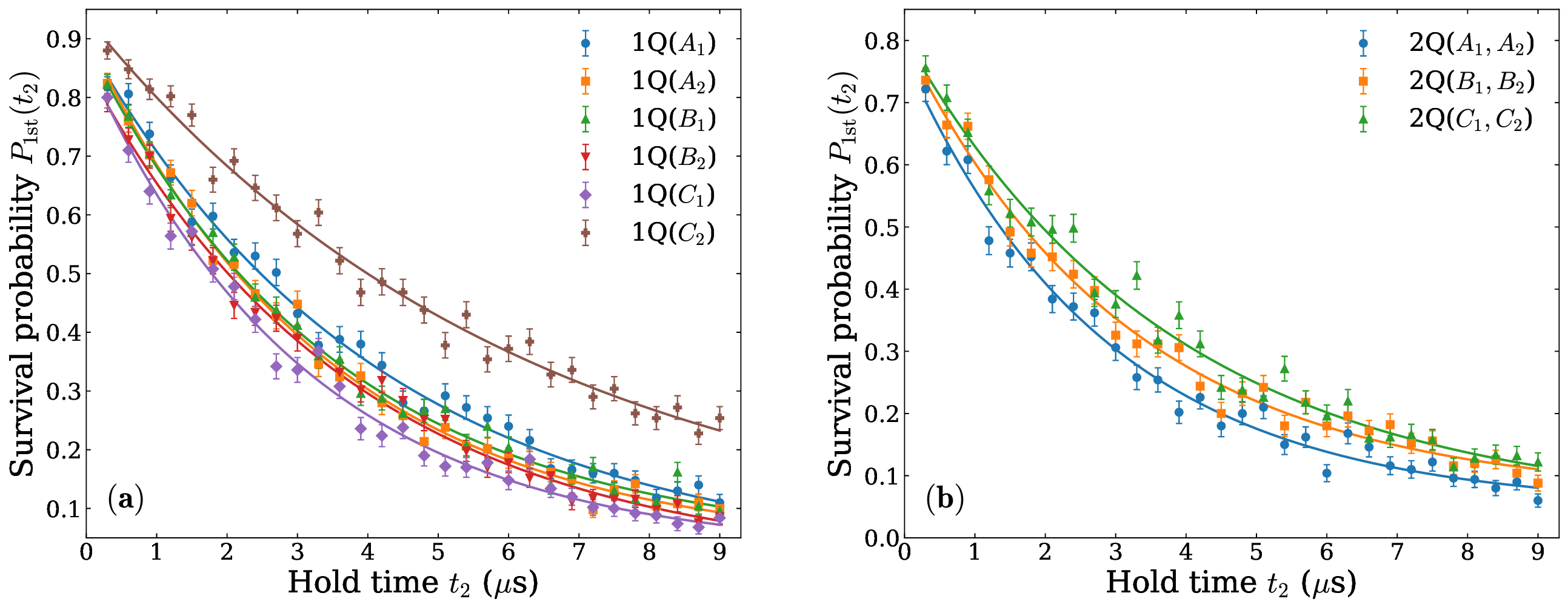}
\caption{
Representative energy-relaxation curves for the single-qubit and interacting two-qubit systems.
For all measurement conditions listed in Table~\ref{tab:measurement_params}, the survival probability was measured as a function of the hold time $t_2$, and the energy relaxation rate was extracted using the same weighted exponential fitting procedure.
Panel (a) shows a representative single-qubit result corresponding to measurement condition No.~5 in Table~\ref{tab:measurement_params}(a), whereas panel (b) shows a representative interacting two-qubit result corresponding to measurement condition No.~9 in Table~\ref{tab:measurement_params}(b).
The symbols represent the experimentally measured survival probabilities, and the solid curves represent the weighted exponential fits.
The error bars indicate the binomial standard errors of the measured survival probabilities.
}
\label{fig:survival_probability_fit}
\end{figure*}
Fig.~\ref{fig:survival_probability_fit} shows representative energy-relaxation curves for the single-qubit and interacting two-qubit systems. The single-qubit result corresponds to No.~5 in Table~\ref{tab:measurement_params}(a), whereas the interacting two-qubit result corresponds to No.~9 in Table~\ref{tab:measurement_params}(b). The symbols represent the experimentally measured survival probabilities, and the error bars represent the binomial standard errors of the measured survival probabilities.

For each hold time $t_2$, the reverse-annealing protocol was repeated for a fixed number of shots. We counted how many times the spin configuration corresponding to the first excited state was observed after the reverse-annealing sequence. The fraction of these observations among the total number of shots was defined as the survival probability $P_{1\mathrm{st}}(t_2)$.
The measured survival probabilities were fitted using the following exponential function, taking the binomial standard error of each data point into account:
\begin{equation}
f(t_2)=
p_0
\exp\left(-\frac{t_2}{T_1}\right)
+
p_\infty,
\label{eq:survival_fit}
\end{equation}
where $p_0$ is the amplitude of the decaying component and $p_\infty$ is the asymptotic offset. The solid curves in Fig.~\ref{fig:survival_probability_fit} represent the fitted results. The obtained $T_1$ was taken as the energy relaxation time, and the corresponding energy relaxation rate was calculated as
\begin{equation}
\Gamma_{\mathrm{exp}}=\frac{1}{T_1}.
\label{eq:relaxation_rate_exp}
\end{equation}
The uncertainty in $T_1$ was obtained from the exponential fit, and the corresponding uncertainty in $\Gamma_{\mathrm{exp}}$ was calculated from the uncertainty in $T_1$.

\subsection{Single-qubit relaxation rates and squared transition matrix elements}
\begin{figure}[htbp]
    \centering
    \includegraphics[width=1\linewidth]{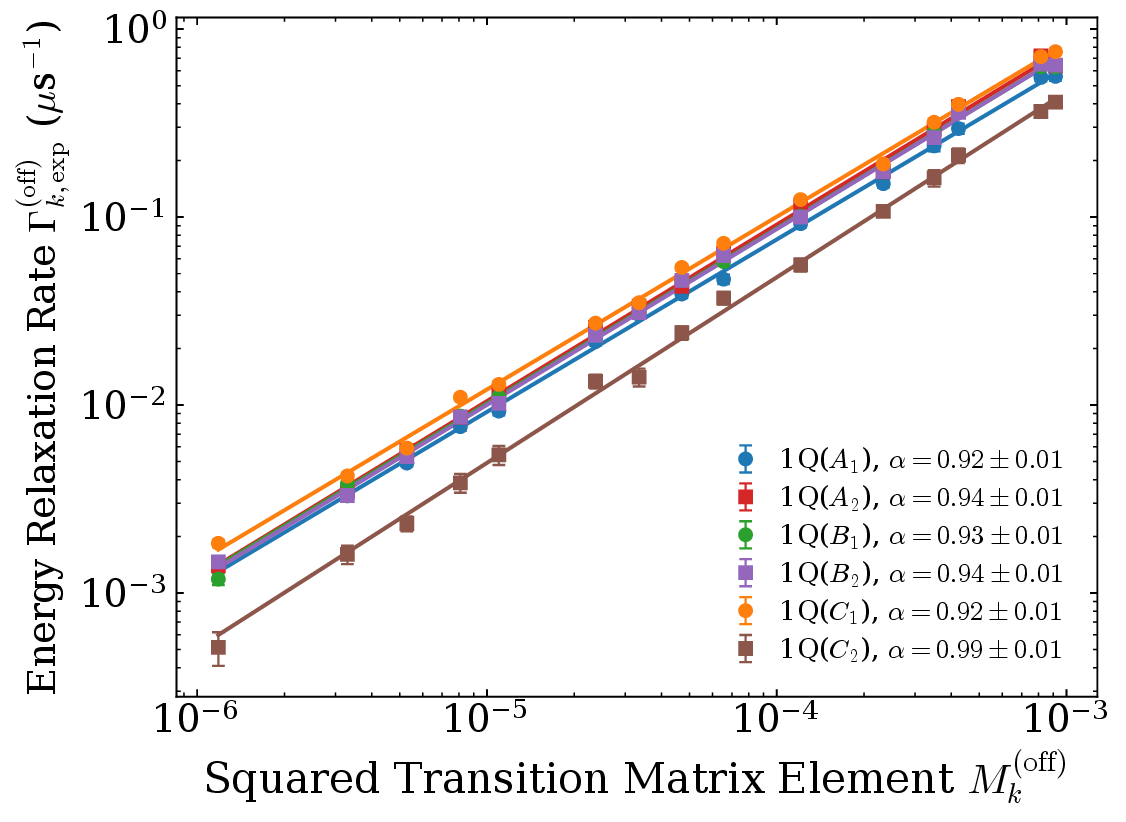}
    \caption{Relationship between the squared transition matrix elements and the experimentally measured energy relaxation rates for single qubits in the non-interacting case.
All data points shown in this figure correspond to the non-interacting measurement conditions listed in Table~\ref{tab:measurement_params}(a).
Each point represents the result obtained by measuring each physical qubit as an individual single qubit.
The horizontal axis shows the squared transition matrix element of qubit $k$, $M^{(\mathrm{off})}_{k}$, and the vertical axis shows the corresponding experimentally measured energy relaxation rate, $\Gamma^{(\mathrm{off})}_{k,\mathrm{exp}}$.
Here, $k=1,2$ denotes the qubit index within each pair.
For example, 1Q(A1) and 1Q(A2) in the legend correspond to the first and second qubits in pair A, respectively.
The same applies to pairs B and C.
The quantity $\alpha$ shown in the figure represents the scaling exponent obtained from fitting all data points for each physical qubit separately using Eq.~(\ref{eq:fit_function}).
}
    \label{fig:1Q_Gamma_vs_M}
\end{figure}

Fig.~\ref{fig:1Q_Gamma_vs_M} shows the relationship between the squared transition matrix elements of the single qubits in the non-interacting system and the experimentally measured energy relaxation rates, $\Gamma^{(\mathrm{off})}_{1,\mathrm{exp}}$ and $\Gamma^{(\mathrm{off})}_{2,\mathrm{exp}}$.

To evaluate the correlation between the experimentally measured single-qubit energy relaxation rates and the squared transition matrix elements, we fitted the experimental data using the following power-law form:
\begin{equation}
\Gamma^{(\mathrm{off})}_{k,\mathrm{exp}}
=
a\left(M^{(\mathrm{off})}_k\right)^{\alpha},
\qquad k=1,2,
\label{eq:fit_function}
\end{equation}
where $a$ is a fitting parameter and $\alpha$ is the scaling exponent.

Under the assumption of independent $\sigma^z_k$-type noise and approximately constant effective transition-rate coefficients $\gamma_k(\Delta E)$ over the measurement conditions, the model predicts a linear dependence between the energy relaxation rate and the squared transition matrix element, corresponding to $\alpha=1$. We therefore use the fitted value of $\alpha$ to evaluate the extent to which the experimentally observed relaxation behavior follows the scaling predicted by the model.
The resulting scaling exponent for each qubit was approximately $\alpha=0.92$--$0.99$. This result indicates that, over the investigated parameter range, the single-qubit relaxation rates approximately follow the linear scaling predicted by the model adopted in this study.

\subsection{Two-qubit relaxation rates and squared transition matrix elements}
\begin{figure}[htbp]
\centering
\includegraphics[
width=\linewidth
]{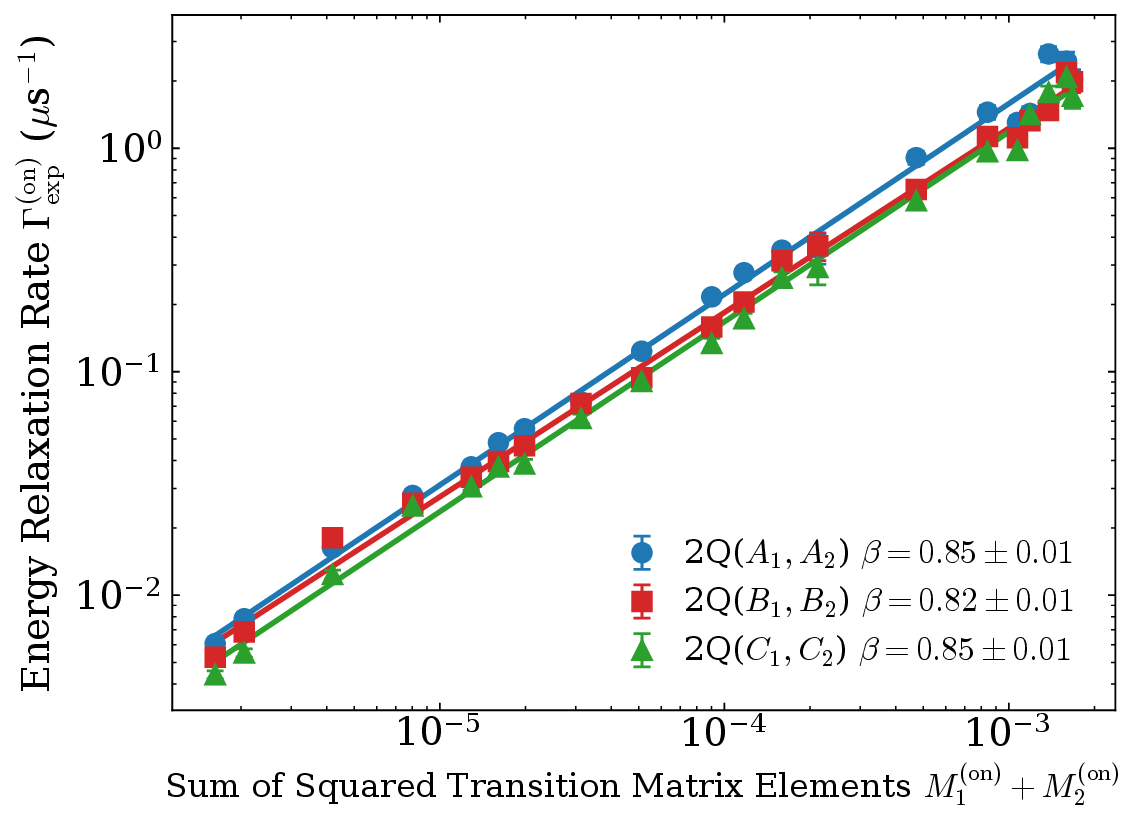}
\caption{
Relationship between the sum of the squared transition matrix elements
and the experimentally measured energy relaxation rate for the
interacting two-qubit system.
All data points shown in this figure correspond to the interacting
two-qubit measurement conditions listed in Table~\ref{tab:measurement_params}(b).
The horizontal axis shows
$M^{(\mathrm{on})}_1+M^{(\mathrm{on})}_2$,
and the vertical axis shows
$\Gamma^{(\mathrm{on})}_{\mathrm{exp}}$.
The value of $\beta$ represents the scaling exponent obtained by
fitting all data points for each qubit pair separately using
Eq.~(\ref{eq:fit_function_2Q}).
The obtained exponents are below the linear scaling
$\beta=1$ predicted by the model adopted in this study,
which assumes independent $\sigma^z$-type noise and approximately
constant transition-rate coefficients $\gamma_k(\Delta E)$.}
\label{fig:2Q_Gamma_vs_M}
\end{figure}

Fig.~\ref{fig:2Q_Gamma_vs_M} shows the relationship between the sum of the squared transition matrix elements,
$M^{(\mathrm{on})}_1+M^{(\mathrm{on})}_2$,
for the interacting two-qubit system and the experimentally measured energy relaxation rate,
$\Gamma^{(\mathrm{on})}_{\mathrm{exp}}$.

To examine how the experimentally measured energy relaxation rate depends on the sum of the squared transition matrix elements, we fitted the experimental data using the following power-law form:
\begin{equation}
\Gamma^{(\mathrm{on})}_{\mathrm{exp}}
=
b\left(
M^{(\mathrm{on})}_1+M^{(\mathrm{on})}_2
\right)^{\beta},
\label{eq:fit_function_2Q}
\end{equation}
where $b$ is a fitting parameter and $\beta$ is the scaling exponent.

Under the assumption of independent $\sigma^z_k$-type noise and approximately constant effective transition-rate coefficients $\gamma_k(\Delta E)$ over the measurement conditions, the model predicts a linear dependence between the energy relaxation rate and the sum of the squared transition matrix elements, corresponding to $\beta=1$.
The resulting scaling exponent for each qubit pair was approximately $\beta=0.82$--$0.85$. The fitted values are below the linear-scaling value $\beta=1$, indicating a deviation from the linear dependence expected from the independent $\sigma^z_k$-type noise model assumed in this study.
Appendix~\ref{app:concurrence} further characterizes the concurrence of the ideal two-qubit eigenstates and
examines how the concurrence and the local transition
matrix elements co-vary along the parameter path
investigated here.

\subsection{Comparison between predicted and measured two-qubit relaxation rates}

We next predicted the energy relaxation rates of the interacting
two-qubit systems using the experimentally measured single-qubit
relaxation rates in the non-interacting systems.

\begin{figure*}[t]
\centering
\includegraphics[
width=0.95\textwidth
]{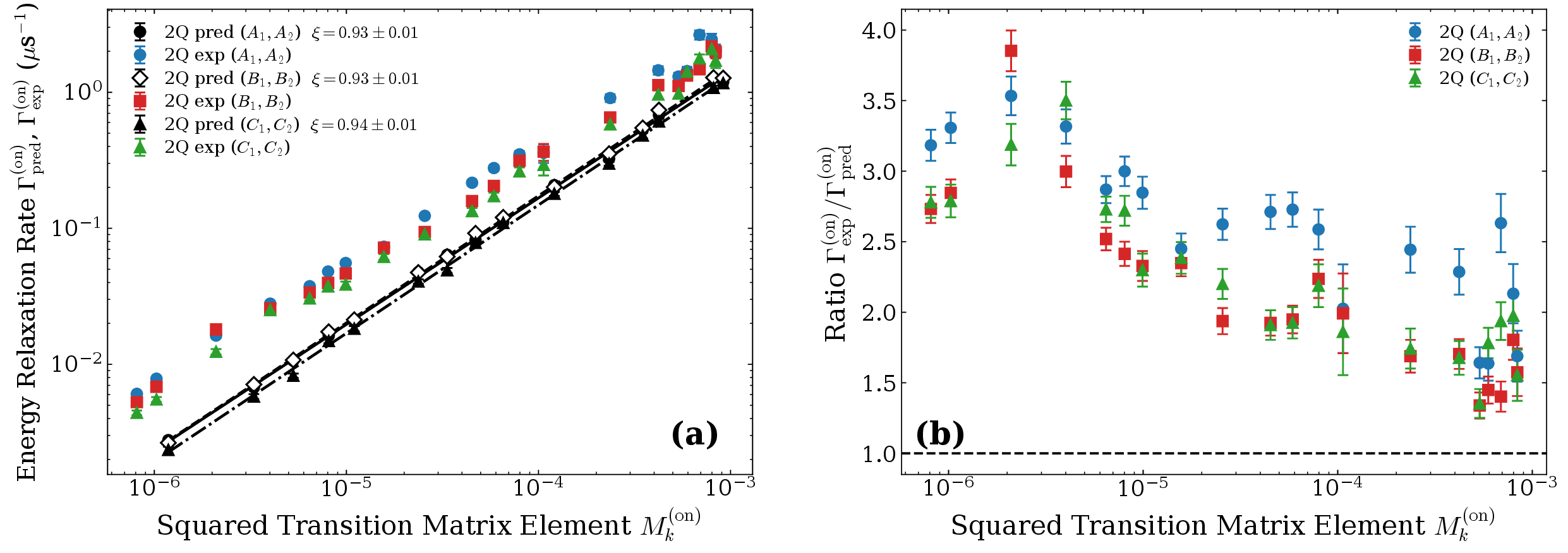}
\caption{
Comparison between the predicted and experimentally measured energy
relaxation rates for the interacting two-qubit system.
The predictions were obtained using the non-interacting single-qubit
measurement data listed in Table~\ref{tab:measurement_params}(a),
whereas the experimentally measured two-qubit relaxation rates correspond
to the interacting measurement conditions listed in
Table~\ref{tab:measurement_params}(b).
Panel (a) shows the predicted energy relaxation rates,
$\Gamma^{(\mathrm{on})}_{\mathrm{pred}}$,
derived from the experimentally measured single-qubit relaxation rates
$\Gamma^{(\mathrm{off})}_{1,\mathrm{exp}}$ and
$\Gamma^{(\mathrm{off})}_{2,\mathrm{exp}}$,
together with the experimentally measured two-qubit relaxation rates,
$\Gamma^{(\mathrm{on})}_{\mathrm{exp}}$.
The horizontal axis shows the squared transition matrix element
$M^{(\mathrm{on})}_k$,
and the vertical axis shows the energy relaxation rate.
Panel (b) shows the ratio
$\Gamma^{(\mathrm{on})}_{\mathrm{exp}}/
\Gamma^{(\mathrm{on})}_{\mathrm{pred}}$
as a function of $M^{(\mathrm{on})}_k$.
The dashed horizontal line represents a ratio of unity.
}
\label{fig:2Q_pred_exp_and_ratio}
\end{figure*}

For each qubit pair, the experimentally measured relaxation rates of
the two constituent single qubits obtained under the same measurement
condition were summed. The summed relaxation rates were then fitted as
a function of $M_k^{(\mathrm{off})}$ using the power-law relation
\begin{equation}
\Gamma_{1,\mathrm{exp}}^{(\mathrm{off})}
+
\Gamma_{2,\mathrm{exp}}^{(\mathrm{off})}
=
c
\left(
M_k^{(\mathrm{off})}
\right)^{\xi},
\label{eq:off_sum_fit}
\end{equation}
where $c$ and $\xi$ are fitting parameters.
The fit was performed independently for each qubit pair.
The fitted values of $\xi$ for each qubit pair are shown in
Fig.~\ref{fig:2Q_pred_exp_and_ratio}(a).
The exponent $\xi$ characterizes the dependence of the summed
single-qubit relaxation rates on the squared transition matrix element.
For constant local relaxation coefficients, the independent
$\sigma^z$-type noise model predicts a linear dependence,
corresponding to $\xi=1$.
Deviations of the fitted exponents from unity therefore indicate
deviations from this linear scaling.

For the symmetric Hamiltonians considered here,
$M_1^{(\mathrm{off})}=M_2^{(\mathrm{off})}\equiv M_k^{(\mathrm{off})}$
and
$M_1^{(\mathrm{on})}=M_2^{(\mathrm{on})}\equiv M_k^{(\mathrm{on})}$.
For each interacting two-qubit measurement condition, we evaluated the fitted sum of
the single-qubit relaxation rates at a reference value of
$M_k^{(\mathrm{off})}$ set equal to the corresponding value of
$M_k^{(\mathrm{on})}$. The resulting value was taken as the
predicted energy relaxation rate
$\Gamma_{\mathrm{pred}}^{(\mathrm{on})}=c
\left(
M_k^{(\mathrm{on})}
\right)^{\xi}$.
This procedure provides an empirical implementation
of the calibration idea underlying Eq.~(\ref{eq:Gamma_on_pred_gamma}), using
matrix-element-matched single-qubit reference rates.
Figure~\ref{fig:2Q_pred_exp_and_ratio}(a) compares the predicted
relaxation rates with the experimentally measured relaxation rates
$\Gamma_{\mathrm{exp}}^{(\mathrm{on})}$.

As shown in Fig.~\ref{fig:2Q_pred_exp_and_ratio}(a), the predicted and experimentally measured energy relaxation rates exhibit similar dependences on the squared transition matrix element.
To quantitatively evaluate their agreement, we calculated the ratio
\begin{equation}
\frac{
\Gamma^{(\mathrm{on})}_{\mathrm{exp}}
}{
\Gamma^{(\mathrm{on})}_{\mathrm{pred}}
}.
\end{equation}
The dependence of this ratio on the squared transition matrix element is shown in Fig.~\ref{fig:2Q_pred_exp_and_ratio}(b).
In the region where the squared transition matrix element
$M^{(\mathrm{on})}_k$ is large, the ratio decreases
toward unity, indicating that the discrepancy becomes
smaller, although the prediction remains below the
measured rate for most data points.
In contrast, as $M^{(\mathrm{on})}_k$ decreases, the ratio tends to increase
to approximately 2--4.
Thus, although the independent $\sigma^z$-type noise model tends to
underestimate the relaxation rate in the small-$M^{(\mathrm{on})}_k$ region,
the predicted and measured values remain within the same
order of magnitude.

One possible origin of this discrepancy is correlated noise acting on the
two qubits.
In the theoretical model used in this study, correlations between the noise
acting on different qubits are neglected, and the noise correlation matrix
is assumed to satisfy
$\gamma_{kl}(\omega)\propto\delta_{kl}$.
Under this assumption, the two-qubit relaxation rate is expressed as the sum
of the independent contributions from the individual qubits.
However, in the presence of correlated noise, cross terms proportional to
$\gamma_{12}(\omega)$ and $\gamma_{21}(\omega)$ can contribute to the
relaxation rate.
These additional contributions may cause the experimentally measured
relaxation rate to exceed the predicted value.

Another possible origin is fluctuations in the qubit--qubit interaction
strength $J_{12}$.
Such fluctuations are specific to the interacting two-qubit system and are
not included in the independent $\sigma^z$-type noise model.
In the small-$M^{(\mathrm{on})}_k$ region, where the relaxation contribution associated with
the independent $\sigma^z$-type noise is strongly suppressed, relaxation induced
by fluctuations in $J_{12}$ may become relatively significant.
However, the detailed studies of such fluctuations are left for future work. 




\section{Conclusion}

In conclusion, we experimentally measured the energy relaxation rates from the first excited state to the ground state in single-qubit systems and an interacting two-qubit system on a D-Wave quantum annealer, and evaluated the applicability of a model based on a GKSL master equation with independent local $\sigma^z$-type noise.

First, we examined the relationship between the experimentally measured single-qubit energy relaxation rates in the non-interacting system, $\Gamma^{(\mathrm{off})}_{1,\mathrm{exp}}$ and $\Gamma^{(\mathrm{off})}_{2,\mathrm{exp}}$, and the corresponding squared transition matrix elements, $M^{(\mathrm{off})}_1$ and $M^{(\mathrm{off})}_2$.
As a result, the energy relaxation rates were found to scale approximately linearly with the squared transition matrix elements, with scaling exponents of $\alpha \simeq 0.92$--$0.99$.
This result supports the approximate
scaling predicted by the assumed noise model, subject to
the statistical uncertainty of the fitted exponent and the
assumed constancy of $\gamma_k$.

Next, using the experimentally measured single-qubit energy relaxation rates as reference values, we predicted the energy relaxation rate of the interacting two-qubit system, $\Gamma^{(\mathrm{on})}_{\mathrm{pred}}$.
In the region where the squared transition matrix elements of the interacting two-qubit system, $M^{(\mathrm{on})}_1$ and $M^{(\mathrm{on})}_2$, are large, the discrepancy between $\Gamma^{(\mathrm{on})}_{\mathrm{pred}}$ and $\Gamma^{(\mathrm{on})}_{\mathrm{exp}}$ is smaller.
On the other hand, in the region where $M^{(\mathrm{on})}_1$ and $M^{(\mathrm{on})}_2$ are small, the measured values exceeded the predicted values by a factor of approximately 2--4, with the prediction remaining within a factor of approximately four over the
investigated range.

These results indicate that the relaxation rate of the investigated
two-qubit eigenstates can be estimated within a factor of
approximately four using the single-qubit relaxation rates obtained in the non-interacting system.
This approach may provide a useful starting point for tests on larger interacting
multiqubit systems.

This work was supported by
JST Moonshot R\&D Grant
Number JPMJMS226C, 
JST CREST Grant Number JPMJCR23I5, and JST PRESTO Grant Number JPMJPR245B.

\appendix \section{Low-Energy Eigenstates, Concurrence,
and Local Transition Matrix Elements of the
Interacting Two-Qubit Hamiltonian} 
\label{app:concurrence}

In this Appendix, we analyze the ground state and first excited state of the ideal interacting two-qubit Hamiltonian introduced in the main text and clarify how state mixing induced by the transverse field is reflected in the entanglement of the eigenstates and in the local $\sigma^z_k$ transition matrix elements. The concurrence evaluated here is a theoretical quantity characterizing the pure eigenstates of the ideal Hamiltonian and is not a direct measurement of the entanglement of the actual open-system state generated during the reverse-annealing process.

In the following, we first obtain the ground state and first excited state using perturbation theory in the weak-transverse-field regime and show that both the local transition matrix element and the concurrence arise at second order in the transverse-field strength. We then examine the range of validity of the perturbative treatment by comparison with exact diagonalization. Finally, we present numerical results along the parameter path used in the experiment and compare how the concurrence, transition matrix element, and experimentally measured relaxation rate vary along this path.

\subsection{Weak-transverse-field analysis of the interacting two-qubit Hamiltonian}
We analyze the low-energy eigenstates of the interacting two-qubit Hamiltonian used in the main text and investigate how state mixing induced by the transverse field affects the local transition matrix element and the concurrence.

In the following, we consider the interacting two-qubit Hamiltonian used in the main text,
\begin{align}
H^{(\mathrm{on})}(s)
=
&-\frac{A(s)}{2}
\left(
\sigma_1^x+\sigma_2^x
\right)\nonumber\\ 
&+\frac{B(s)}{2}
\left[
h
\left(
\sigma_1^z+\sigma_2^z
\right)
+
J_{12}\sigma_1^z\sigma_2^z
\right].
\label{eq:general_two_qubit_hamiltonian}
\end{align}
Here, $\sigma_i^x$ and $\sigma_i^z$ denote the Pauli-$x$ and Pauli-$z$ operators, respectively, acting on the $i$th qubit. The definitions of $A(s)$, $B(s)$, $h$, $J_{12}$, and the annealing parameter $s$ are the same as those used in the main text. As in the main text, we consider the ferromagnetic interaction $J_{12}=-1$.
For the perturbative analysis, we introduce the dimensionless transverse-field parameter
\begin{equation}
b_x
=
-\frac{A(s)}{B(s)}.
\end{equation}
Furthermore, dividing the Hamiltonian by $B(s)/2$ gives
\begin{equation}
\frac{2H^{(\mathrm{on})}(s)}{B(s)}
=
b_x\left(\sigma_1^x+\sigma_2^x\right)
+
h\left(\sigma_1^z+\sigma_2^z\right)
-
\sigma_1^z\sigma_2^z.
\label{eq:dimensionless_two_qubit_hamiltonian}
\end{equation}
For notational simplicity, we denote the normalized Hamiltonian $2H^{(\mathrm{on})}(s)/B(s)$ simply by $H$ in the following.

For the eigenstates of $\sigma^z$, we adopt the convention
\begin{equation}
    \sigma^z\lvert0\rangle=+\lvert0\rangle,
    \qquad
    \sigma^z\lvert1\rangle=-\lvert1\rangle.
\end{equation}
We take the computational basis of the two-qubit system to be
\begin{equation}
    \left\{
        \lvert00\rangle,
        \lvert01\rangle,
        \lvert10\rangle,
        \lvert11\rangle
    \right\}.
\end{equation}

We consider the weak-transverse-field regime satisfying
\begin{equation}
    |b_x|\ll1.
    \label{eq:weak_field_simple_condition}
\end{equation}
Under this assumption, we apply nondegenerate perturbation theory up to second order, treating $b_x$ as the perturbation parameter.

\subsubsection{Perturbed Hamiltonian}
We decompose the Hamiltonian as
\begin{equation}
    H=H_0+V,
\end{equation}
and define the unperturbed Hamiltonian $H_0$ and the perturbation $V$ as
\begin{equation}
    H_0
    =
    h\left(\sigma_1^z+\sigma_2^z\right)
    -
    \sigma_1^z\sigma_2^z,
    \qquad
    V
    =
    b_x\left(\sigma_1^x+\sigma_2^x\right).
    \label{eq:unperturbed_and_perturbation}
\end{equation}
The computational basis states $\lvert n_1n_2\rangle$ are eigenstates of $H_0$, and for $n_1,n_2\in\{0,1\}$, the unperturbed Hamiltonian can be written as
\begin{equation}
H_0
=
\sum_{n_1=0}^{1}
\sum_{n_2=0}^{1}
E_{n_1n_2}^{(0)}
\lvert n_1n_2\rangle
\langle n_1n_2\rvert.
\label{eq:unperturbed_hamiltonian_spectral}
\end{equation}
The corresponding unperturbed eigenenergies are
\begin{align}
    E_{00}^{(0)}&=2h-1,\\
    E_{01}^{(0)}&=1,\\
    E_{10}^{(0)}&=1,\\
    E_{11}^{(0)}&=-2h-1.
    \label{eq:unperturbed_energies}
\end{align}

In the region $0<h<1$,
\begin{equation}
    E_{11}^{(0)}<E_{00}^{(0)}<E_{01}^{(0)}=E_{10}^{(0)}.
\end{equation}
Throughout the perturbative analysis below, we consider fixed values of $h$ in the range $0<h<1$ and take $|b_x|$ sufficiently small that the perturbative corrections remain small. The limits $h \rightarrow 0$ and $h \rightarrow1$, where degeneracies occur, require separate treatment. Therefore, the ground state and first excited state of the unperturbed system are, respectively,
\begin{equation}
    |\Phi^{(\mathrm{on})(0)}_{\mathrm{gs}}\rangle=\lvert11\rangle,
    \qquad
    |\Phi^{(\mathrm{on})(0)}_{\mathrm{1st}}\rangle=\lvert00\rangle.
    \label{eq:unperturbed_ground_and_excited_states}
\end{equation}

Because the perturbation $V$ flips only one qubit, its nonzero matrix elements are
\begin{equation}
    \langle01|V|00\rangle=\langle10|V|00\rangle=\langle01|V|11\rangle=\langle10|V|11\rangle=b_x,
    \label{eq:nonzero_perturbation_matrix_elements}
\end{equation}
together with their Hermitian conjugates.

\subsubsection{State coupling induced by the transverse field}
To clarify the structure of the transverse-field-induced coupling between the states, we introduce the symmetric and antisymmetric states under qubit exchange,
\begin{equation}
    |S\rangle
    =
    \frac{|01\rangle+|10\rangle}{\sqrt{2}},
    \qquad
    |A\rangle
    =
    \frac{|01\rangle-|10\rangle}{\sqrt{2}}.
    \label{eq:symmetric_antisymmetric_states}
\end{equation}
Using the basis
\begin{equation}
    \left\{
    |11\rangle,
    |S\rangle,
    |00\rangle,
    |A\rangle
    \right\},
\end{equation}
the Hamiltonian can be written as
\begin{equation}
H
=
\begin{pmatrix}
-1-2h & \sqrt{2}b_x & 0 & 0\\
\sqrt{2}b_x & 1 & \sqrt{2}b_x & 0\\
0 & \sqrt{2}b_x & -1+2h & 0\\
0 & 0 & 0 & 1
\end{pmatrix}.
\label{eq:hamiltonian_symmetric_basis}
\end{equation}
Because the transverse-field term $b_x(\sigma_1^x+\sigma_2^x)$ is invariant under qubit exchange, it does not mix the symmetric and antisymmetric subspaces. Therefore, the antisymmetric state $|A\rangle$ is decoupled from the other three states. In contrast, within the symmetric subspace, the states are coupled according to
\begin{equation}
    |11\rangle
    \longleftrightarrow
    |S\rangle
    \longleftrightarrow
    |00\rangle.
    \label{eq:state_mixing_path}
\end{equation}
Since a single application of the transverse-field term flips only one qubit, $|11\rangle$ and $|00\rangle$ are not directly coupled but are effectively coupled through a second-order perturbative process mediated by $|S\rangle$.

\subsubsection{Ground state from perturbation theory}

The energy difference between the unperturbed ground state $\lvert11\rangle$ and the states obtained by flipping one qubit from $\lvert11\rangle$ is
\begin{equation}
    E_{11}^{(0)}-E_{01}^{(0)}
    =
    E_{11}^{(0)}-E_{10}^{(0)}
    =
    -2(1+h).
\end{equation}
Therefore, the first-order correction to the ground state is
\begin{align}
    \sum_{m= 00, 01, 10}
    \frac{
        \langle m|V|11\rangle
    }{
        E_{11}^{(0)}-E_m^{(0)}
    }
    \lvert m\rangle \nonumber\\
    =
    -\frac{b_x}{2(1+h)}
    \left(
        \lvert01\rangle+\lvert10\rangle
    \right).
    \label{eq:first_order_ground_state}
\end{align}

The transition from $\lvert11\rangle$ to $\lvert00\rangle$ requires two qubit flips, with $\lvert01\rangle$ and $\lvert10\rangle$ contributing as intermediate states. Therefore, the second-order contribution to the ground state along $\lvert00\rangle$ is
\begin{align}
    &\sum_{m=01,10}
    \frac{
        \langle00|V|m\rangle
        \langle m|V|11\rangle
    }{
        \left(
            E_{11}^{(0)}-E_{00}^{(0)}
        \right)
        \left(
            E_{11}^{(0)}-E_m^{(0)}
        \right)
    }|00\rangle \nonumber\\
    &=
    \frac{b_x^2}{4h(1+h)}|00\rangle.
    \label{eq:second_order_ground_00_coefficient}
\end{align}

Thus, the normalized ground state up to second order in $b_x$ is
\begin{align}
    |\Phi^{(\mathrm{on})}_{\mathrm{gs}}\rangle
    =&\left[1-\frac{b_x^2}{4(1+h)^2}\right]\lvert11\rangle \nonumber\\
    &-\frac{b_x}{2(1+h)}\left(\lvert01\rangle+\lvert10\rangle\right)\nonumber\\
    &+\frac{b_x^2}{4h(1+h)}\lvert00\rangle+O(b_x^3).
    \label{eq:perturbative_ground_state}
\end{align}

\subsubsection{First excited state from perturbation theory}

The energy difference between the unperturbed first excited state $\lvert00\rangle$ and the states obtained by flipping one qubit from $\lvert00\rangle$ is
\begin{equation}
    E_{00}^{(0)}-E_{01}^{(0)}
    =
    E_{00}^{(0)}-E_{10}^{(0)}
    =
    -2(1-h).
\end{equation}
Therefore, the first-order correction to the first excited state is
\begin{align}
    |\Phi^{(\mathrm{on})(1)}_{\mathrm{1st}}\rangle
    &=
    \sum_{m\neq00}
    \frac{
        \langle m|V|00\rangle
    }{
        E_{00}^{(0)}-E_m^{(0)}
    }
    \lvert m\rangle\nonumber\\
    &=-\frac{b_x}{2(1-h)}
    \left(
        \lvert01\rangle+\lvert10\rangle
    \right).
    \label{eq:first_order_excited_state}
\end{align}

The transition from $\lvert00\rangle$ to $\lvert11\rangle$ requires two qubit flips, with $\lvert01\rangle$ and $\lvert10\rangle$ contributing as intermediate states. Therefore, the second-order contribution to the first excited state along $\lvert11\rangle$ is
\begin{align}
    &\sum_{m=01,10}
    \frac{
        \langle11|V|m\rangle
        \langle m|V|00\rangle
    }{
        \left(
            E_{00}^{(0)}-E_{11}^{(0)}
        \right)
        \left(
            E_{00}^{(0)}-E_m^{(0)}
        \right)
    }|11\rangle \nonumber\\
    &=
    -\frac{b_x^2}{4h(1-h)}|11\rangle.
    \label{eq:second_order_excited_11_coefficient}
\end{align}

Thus, the normalized first excited state up to second order in $b_x$ is
\begin{align}
    |\Phi^{(\mathrm{on})}_{\mathrm{1st}}\rangle
    =&
    \left[
        1-\frac{b_x^2}{4(1-h)^2}
    \right]\lvert00\rangle\nonumber\\
    &-\frac{b_x}{2(1-h)}
    \left(
        \lvert01\rangle+\lvert10\rangle
    \right)\nonumber\\
    &-
    \frac{b_x^2}{4h(1-h)}
    \lvert11\rangle
    +O(b_x^3).
    \label{eq:perturbative_excited_state}
\end{align}

\subsubsection{Transition matrix element}

We define the magnitude of the local transition matrix element between the ground state and first excited state for the $k$th qubit as
\begin{equation}
\mu_k^{(\mathrm{on})}
=
\left|
\langle
\Phi_{\mathrm{gs}}^{(\mathrm{on})}
|
\sigma_k^z
|
\Phi_{\mathrm{1st}}^{(\mathrm{on})}
\rangle
\right|,
\end{equation}
and its square as
\begin{equation}
M_k^{(\mathrm{on})}
=
\left(
\mu_k^{(\mathrm{on})}
\right)^2.
\end{equation}

Since this Hamiltonian is symmetric under qubit exchange, we explicitly consider the case $k=1$ below.

The action of $\sigma_1^z$ on the computational basis states is
\begin{align}
    \sigma_1^z\lvert00\rangle
    &=
    +\lvert00\rangle,
    \\
    \sigma_1^z\lvert01\rangle
    &=
    +\lvert01\rangle,
    \\
    \sigma_1^z\lvert10\rangle
    &=
    -\lvert10\rangle,
    \\
    \sigma_1^z\lvert11\rangle
    &=
    -\lvert11\rangle.
\end{align}

For the symmetric and antisymmetric states defined in Eq.~\eqref{eq:symmetric_antisymmetric_states}, the action of $\sigma_1^z$ gives
\begin{equation}
    \sigma_1^z|S\rangle
    =
    |A\rangle,
    \qquad
    \sigma_1^z|A\rangle
    =
    |S\rangle,
\end{equation}
and hence
\begin{equation}
    \langle S|\sigma_1^z|S\rangle
    =
    \langle S|A\rangle
    =
    0.
\end{equation}
The Hamiltonian is symmetric under qubit exchange, and both the ground state and first excited state belong to the symmetric subspace and contain no component of the antisymmetric state $|A\rangle$. Therefore, the $|S\rangle$ components generated by first-order perturbation do not contribute to the transition matrix element of $\sigma_1^z$. This implies that the lowest-order nonzero contribution to the transition matrix element is $O(b_x^2)$ rather than $O(b_x)$. In fact, the second-order contribution arising from the product of the first-order $\lvert01\rangle$ and $\lvert10\rangle$ components cancels because these states have opposite eigenvalues of $\sigma_1^z$,
\begin{align}
    &\left[-\frac{b_x}{2(1-h)}\right]\left[-\frac{b_x}{2(1+h)}\right]\langle01|\sigma_1^z|01\rangle \nonumber\\
    &+ \left[-\frac{b_x}{2(1-h)}\right]\left[-\frac{b_x}{2(1+h)}\right]\langle10|\sigma_1^z|10\rangle=0.
\end{align}

Therefore, only the contributions involving the $\lvert00\rangle$ and $\lvert11\rangle$ components remain, yielding
\begin{align}
    \langle\Phi^{(\mathrm{on})}_{\mathrm{gs}}|\sigma_1^z|\Phi^{(\mathrm{on})}_{\mathrm{1st}}\rangle
    &=
    \frac{b_x^2}{4h(1+h)}
    +
    \frac{b_x^2}{4h(1-h)}
    +O(b_x^4)
    \nonumber\\
    &=
    \frac{b_x^2}{2h(1-h^2)}
    +O(b_x^4).
    \label{eq:transition_matrix_element_without_absolute}
\end{align}
For $0<h<1$, the leading term on the right-hand side is positive, and therefore
\begin{equation}
    \mu_1^{(\mathrm{on})}
    =
    \frac{b_x^2}{2h(1-h^2)}
    +O(b_x^4).
    \label{eq:transition_matrix_element}
\end{equation}

Thus, the local transition matrix element arises at order $\mu^{(\mathrm{on})}_1=O(b_x^2)$ in the weak-transverse-field regime, while its square, which enters directly into the relaxation rate in the main text, scales as $M^{(\mathrm{on})}_1=(\mu^{(\mathrm{on})}_1)^2=O(b_x^4)$.

\subsubsection{Concurrence}
The concurrence evaluated below characterizes the pure eigenstates of the ideal Hamiltonian obtained above. Note that the open-system state generated during the actual reverse-annealing process is generally a mixed state, and the concurrence considered here does not directly represent the entanglement of the actual open-system state.

Following Wootters~\cite{Wootters1998}, we define the spin-flipped state of a two-qubit pure state $\lvert\psi\rangle$ as
\begin{equation}
    \lvert\widetilde{\psi}\rangle
    =
    \left(\sigma^y\otimes\sigma^y\right)
    \lvert\psi^*\rangle.
    \label{eq:spin_flipped_state}
\end{equation}
Here, $\lvert\psi^*\rangle$ denotes the complex conjugate of $\lvert\psi\rangle$ in the computational basis, and $\sigma^y$ is the Pauli-$y$ operator
\begin{equation}
    \sigma^y
    =
    \begin{pmatrix}
        0 & -i \\
        i & 0
    \end{pmatrix}.
\end{equation}

The concurrence of a two-qubit pure state is defined by
\begin{equation}
C(\psi)
=
\left|
\langle\psi|\widetilde{\psi}\rangle
\right|.
\label{eq:pure_state_concurrence}
\end{equation}

Consider a general normalized two-qubit pure state,
\begin{equation}
\lvert\psi\rangle
=
a_{00}\lvert00\rangle
+
a_{01}\lvert01\rangle
+
a_{10}\lvert10\rangle
+
a_{11}\lvert11\rangle.
\label{eq:general_two_qubit_pure_state}
\end{equation}

Using the action of $\sigma^y$,
\begin{equation}
    \sigma^y\lvert0\rangle
    =
    i\lvert1\rangle,
    \qquad
    \sigma^y\lvert1\rangle
    =
    -i\lvert0\rangle,
\end{equation}
we obtain
\begin{align}
    \left(\sigma^y\otimes\sigma^y\right)\lvert00\rangle
    &=
    -\lvert11\rangle,
    \\
    \left(\sigma^y\otimes\sigma^y\right)\lvert01\rangle
    &=
    \lvert10\rangle,
    \\
    \left(\sigma^y\otimes\sigma^y\right)\lvert10\rangle
    &=
    \lvert01\rangle,
    \\
    \left(\sigma^y\otimes\sigma^y\right)\lvert11\rangle
    &=
    -\lvert00\rangle.
\end{align}
Therefore, the spin-flipped state is
\begin{align}
\lvert\widetilde{\psi}\rangle
={}&
-a_{11}^*\lvert00\rangle
+
a_{10}^*\lvert01\rangle
+
a_{01}^*\lvert10\rangle
-
a_{00}^*\lvert11\rangle.
\label{eq:general_spin_flipped_state}
\end{align}

It follows that
\begin{align}
\langle\psi|\widetilde{\psi}\rangle
&=
-a_{00}^*a_{11}^*
+
a_{01}^*a_{10}^*
+
a_{10}^*a_{01}^*
-
a_{11}^*a_{00}^*
\nonumber\\
&=
2
\left(
a_{01}a_{10}-a_{00}a_{11}
\right)^*.
\end{align}
Thus, the concurrence of a general two-qubit pure state can be written as
\begin{equation}
C(\psi)
=
2
\left|
a_{00}a_{11}-a_{01}a_{10}
\right|.
\label{eq:concurrence_from_coefficients}
\end{equation}

\subsubsection{Concurrence of the ground state}

We define the concurrence of the ground state $|\Phi_{gs}^{(\mathrm{on})}\rangle$ as
\begin{equation}
C_{\mathrm{gs}}^{(\mathrm{on})}
=
C(\Phi_{\mathrm{gs}}^{(\mathrm{on})})
=
\left|
\langle \Phi_{\mathrm{gs}}^{(\mathrm{on})}|\widetilde{\Phi}_{\mathrm{gs}}^{(\mathrm{on})}\rangle
\right|.
\end{equation}

For the ground state obtained by second-order perturbation theory,
\begin{align}
\lvert \Phi_{\mathrm{gs}}^{(\mathrm{on})}\rangle
=&
\left[
1-\frac{b_x^2}{4(1+h)^2}
\right]\lvert11\rangle\nonumber\\
&-\frac{b_x}{2(1+h)}
\left(
\lvert01\rangle+\lvert10\rangle
\right)\nonumber\\
&+
\frac{b_x^2}{4h(1+h)}
\lvert00\rangle+O(b_x^3),
\end{align}
the coefficients are
\begin{align}
a_{00}^{(\mathrm{gs})}
&=
\frac{b_x^2}{4h(1+h)}
+O(b_x^4),
\\
a_{01}^{(\mathrm{gs})}
=
a_{10}^{(\mathrm{gs})}
&=
-\frac{b_x}{2(1+h)}
+O(b_x^3),
\\
a_{11}^{(\mathrm{gs})}
&=
1-\frac{b_x^2}{4(1+h)^2}
+O(b_x^4).
\end{align}

Substituting these coefficients into Eq.~\eqref{eq:concurrence_from_coefficients}, we obtain
\begin{align}
C_{\mathrm{gs}}^{(\mathrm{on})}
&=
2
\left|
a_{00}^{(\mathrm{gs})}a_{11}^{(\mathrm{gs})}
-
a_{01}^{(\mathrm{gs})}a_{10}^{(\mathrm{gs})}
\right|
\nonumber\\
&=
2
\left|
\frac{b_x^2}{4h(1+h)}
-
\frac{b_x^2}{4(1+h)^2}
\right|
+O(b_x^4)
\nonumber\\
&=
\frac{b_x^2}{2h(1+h)^2}
+O(b_x^4).
\label{eq:ground_state_concurrence}
\end{align}

\subsubsection{Concurrence of the first excited state}

Similarly, we define the concurrence of the first excited state $|\Phi_{\mathrm{1st}}^{(\mathrm{on})}\rangle$ as
\begin{equation}
C_{\mathrm{1st}}^{(\mathrm{on})}
=
C(\Phi_{\mathrm{1st}}^{(\mathrm{on})})
=
\left|
\langle \Phi_{\mathrm{1st}}^{(\mathrm{on})}|\widetilde{\Phi}_{\mathrm{1st}}^{(\mathrm{on})}\rangle
\right|.
\end{equation}

For the first excited state obtained by second-order perturbation theory,
\begin{align}
\lvert \Phi_{\mathrm{1st}}^{(\mathrm{on})}\rangle
=&
\left[
1-\frac{b_x^2}{4(1-h)^2}
\right]\lvert00\rangle\nonumber\\
&-\frac{b_x}{2(1-h)}
\left(
\lvert01\rangle+\lvert10\rangle
\right)\nonumber\\
&-
\frac{b_x^2}{4h(1-h)}
\lvert11\rangle
+O(b_x^3),
\end{align}
the coefficients are
\begin{align}
a_{00}^{(\mathrm{1st})}
&=
1-\frac{b_x^2}{4(1-h)^2}
+O(b_x^4),
\\
a_{01}^{(\mathrm{1st})}
=
a_{10}^{(\mathrm{1st})}
&=
-\frac{b_x}{2(1-h)}
+O(b_x^3),
\\
a_{11}^{(\mathrm{1st})}
&=
-\frac{b_x^2}{4h(1-h)}
+O(b_x^4).
\end{align}

Substituting these coefficients into Eq.~\eqref{eq:concurrence_from_coefficients}, we obtain
\begin{align}
C_{\mathrm{1st}}^{(\mathrm{on})}
&=
2
\left|
a_{00}^{(\mathrm{1st})}a_{11}^{(\mathrm{1st})}
-
a_{01}^{(\mathrm{1st})}a_{10}^{(\mathrm{1st})}
\right|
\nonumber\\
&=
2
\left|
-\frac{b_x^2}{4h(1-h)}
-
\frac{b_x^2}{4(1-h)^2}
\right|
+O(b_x^4)
\nonumber\\
&=
\frac{b_x^2}{2h(1-h)^2}
+O(b_x^4).
\label{eq:first_excited_state_concurrence}
\end{align}

Therefore, the concurrences of the ground state and first excited state are, respectively,
\begin{align}
&C_{\mathrm{gs}}^{(\mathrm{on})}
=
\frac{b_x^2}{2h(1+h)^2}
+O(b_x^4),\nonumber\\
&
C_{\mathrm{1st}}^{(\mathrm{on})}
=
\frac{b_x^2}{2h(1-h)^2}
+O(b_x^4).
\label{eq:ground_excited_concurrences}
\end{align}

\subsubsection{Relation between concurrence and the transition matrix element at a fixed longitudinal field}
From Eqs.~\eqref{eq:transition_matrix_element} and \eqref{eq:ground_state_concurrence}, the ratio of the ground-state concurrence to the transition matrix element is
\begin{align}
\frac{C_{\mathrm{gs}}^{(\mathrm{on})}}{\mu_1^{(\mathrm{on})}}
&=
\left[]
\frac{b_x^2}{2h(1+h)^2}
\right]
\left[\frac{b_x^2}{2h(1-h^2)}\right]^{-1}
+O(b_x^2)\nonumber\\
&=
\frac{1-h}{1+h}
+O(b_x^2).
\end{align}
Therefore,
\begin{equation}
C_{\mathrm{gs}}^{(\mathrm{on})}
=
\frac{1-h}{1+h}\mu_1^{(\mathrm{on})}
+O(b_x^4).
\label{eq:ground_concurrence_transition_relation}
\end{equation}

Similarly, for the concurrence of the first excited state,
\begin{align}
\frac{C_{\mathrm{1st}}^{(\mathrm{on})}}{\mu_1^{(\mathrm{on})}}
&=
\left[\frac{b_x^2}{2h(1-h)^2}\right]
\left[\frac{b_x^2}{2h(1-h^2)}\right]^{-1}
+O(b_x^2)\nonumber\\
&=
\frac{1+h}{1-h}
+O(b_x^2).
\end{align}
Therefore,
\begin{equation}
C_{\mathrm{1st}}^{(\mathrm{on})}
=
\frac{1+h}{1-h}\mu_1^{(\mathrm{on})}
+O(b_x^4).
\label{eq:excited_concurrence_transition_relation}
\end{equation}

Thus, along a transverse-field path in which only $b_x$ is varied while $h$ is fixed, the concurrence and the local transition matrix element are proportional at leading order. However, because the proportionality coefficient depends explicitly on $h$, this result does not imply a general parameter-independent relation between concurrence and the transition matrix element.

\begin{figure*}[t]
    \centering
    \includegraphics[
        width=\textwidth
    ]{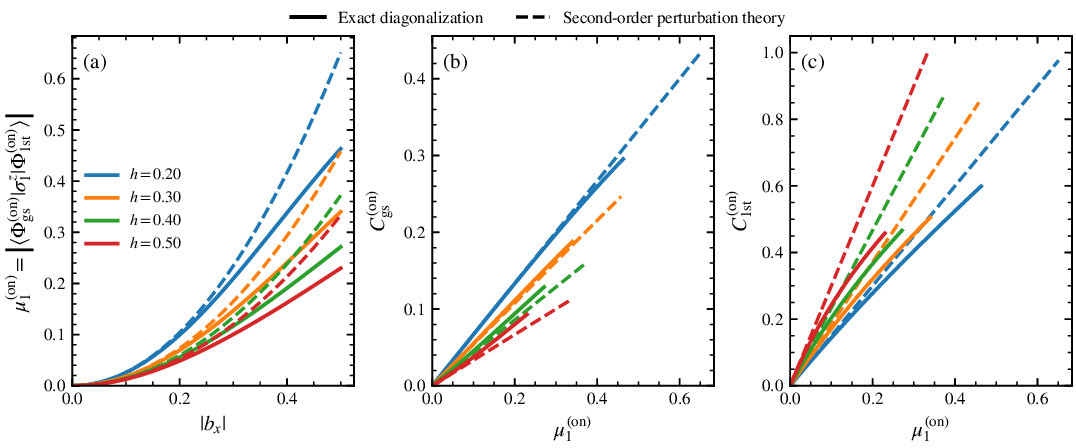}
    \caption{
    Comparison between exact diagonalization and second-order perturbation theory when the transverse field $b_x$ is swept for fixed longitudinal fields $h$.
    The calculations were performed for $h=0.20,\,0.30,\,0.40,\,0.50$ over the range $0\leq |b_x|\leq0.5$. Here, $b_x=-A(s)/B(s)$ is the dimensionless transverse-field parameter, and $h$ is the longitudinal-field parameter used in the main text.
    (a) Transition matrix element $\mu_1^{(\mathrm{on})}$ as a function of $|b_x|$.
    (b) Relation between the ground-state concurrence $C_{\mathrm{gs}}^{(\mathrm{on})}$ and the transition matrix element $\mu_1^{(\mathrm{on})}$.
    (c) Relation between the first-excited-state concurrence $C_{\mathrm{1st}}^{(\mathrm{on})}$ and the transition matrix element $\mu_1^{(\mathrm{on})}$.
    Solid lines represent the results of exact diagonalization, and dashed lines represent the results of second-order perturbation theory.
    }
    \label{fig:fixed_h_B_sweep_perturbation_comparison}
\end{figure*}

To numerically examine the validity of the second-order perturbative expressions derived above, we swept the transverse field $b_x$ for several fixed values of $h$ and compared the results obtained by exact diagonalization with those obtained from second-order perturbation theory. The results are shown in Fig.~\ref{fig:fixed_h_B_sweep_perturbation_comparison}.

As shown in Fig.~\ref{fig:fixed_h_B_sweep_perturbation_comparison}, in the sufficiently weak-transverse-field regime, second-order perturbation theory reproduces the exact-diagonalization results well for both the transition matrix element and the concurrence. As $|b_x|$ increases, however, higher-order corrections can no longer be neglected, and the deviation from the second-order perturbative results increases. The perturbative relations between concurrence and the transition matrix element also agree well with the exact-diagonalization results. This comparison therefore confirms numerically that Eqs.~\eqref{eq:ground_concurrence_transition_relation} and \eqref{eq:excited_concurrence_transition_relation} provide the leading-order relations in the weak-transverse-field regime.

We next consider how much the proportionality coefficient between the ground-state concurrence and the transition matrix element changes when $h$ varies over a narrow range. For two values $0<h_1,h_2<1$, we denote the corresponding transition matrix elements and ground-state concurrences by $\mu_1^{(\mathrm{on})}(h_1)$, $\mu_1^{(\mathrm{on})}(h_2)$, $C_{\mathrm{gs}}^{(\mathrm{on})}(h_1)$, and $C_{\mathrm{gs}}^{(\mathrm{on})}(h_2)$. Then,
\begin{equation}
\left|
\frac{C_{\mathrm{gs}}^{(\mathrm{on})}(h_1)}{\mu_1^{(\mathrm{on})}(h_1)}
-\frac{C_{\mathrm{gs}}^{(\mathrm{on})}(h_2)}{\mu_1^{(\mathrm{on})}(h_2)}
\right|
=
\left|
\frac{1-h_1}{1+h_1}
-
\frac{1-h_2}{1+h_2}
+O(b_x^2)
\right|,
\end{equation}
and hence
\begin{align}
\left|
\frac{1-h_1}{1+h_1}
-
\frac{1-h_2}{1+h_2}
\right|
&=
\frac{
2|h_1-h_2|
}{
(1+h_1)(1+h_2)
}\nonumber\\
&\leq
2|h_1-h_2|.
\end{align}
Therefore, in a sufficiently narrow range of $h$ satisfying
\begin{equation}
|h_1-h_2|\ll1,
\end{equation}
the change in the proportionality coefficient between the ground-state concurrence and the transition matrix element is small. Thus, for $0<h<1$, the relations between $C^{(\mathrm{on})}_{\mathrm{gs}}$ and $\mu^{(\mathrm{on})}_{1}$ for different but sufficiently close values of $h$ can be approximately described by lines with nearly the same slope.

Similarly, for the proportionality coefficient of the first excited state, in the range $0<h_1,h_2<0.5$,
\begin{equation}
\left|
\frac{C_{\mathrm{1st}}^{(\mathrm{on})}(h_1)}{\mu_1^{(\mathrm{on})}(h_1)}
-\frac{C_{\mathrm{1st}}^{(\mathrm{on})}(h_2)}{\mu_1^{(\mathrm{on})}(h_2)}
\right|
=
\left|
\frac{1+h_1}{1-h_1}
-
\frac{1+h_2}{1-h_2}
+O(b_x^2)
\right|,
\end{equation}
and
\begin{align}
\left|
\frac{1+h_1}{1-h_1}
-
\frac{1+h_2}{1-h_2}
\right|
&=
\frac{
2|h_1-h_2|
}{
(1-h_1)(1-h_2)
}\nonumber\\
&\leq8|h_1-h_2|.
\end{align}

Therefore, as in the ground-state case, in a sufficiently narrow range of $h$ satisfying
\begin{equation}
|h_1-h_2|\ll1,
\end{equation}
the change in the proportionality coefficient between the first-excited-state concurrence and the transition matrix element is small. Thus, the relations between $C_{\mathrm{1st}}^{(\mathrm{on})}$ and $\mu_1^{(\mathrm{on})}$ for different but sufficiently close values of $h$ can also be approximately described by lines with nearly the same slope.

\begin{figure*}[t]
    \centering
    \includegraphics[
        width=0.85\textwidth
    ]{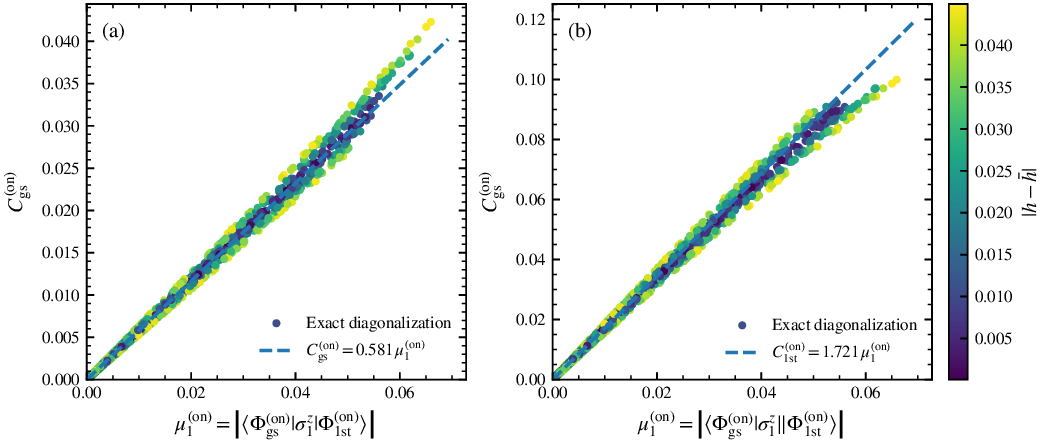}
    \caption{
    Relation between concurrence and the transition matrix element in the weak-transverse-field regime.
    $b_x$ and $h$ were independently sampled from uniform distributions over the ranges $0\leq |b_x|\leq0.17$ and $0.22\leq h\leq0.31$, respectively, with 1000 samples. Each point represents a result obtained by exact diagonalization, and its color indicates the deviation $|h-\bar{h}|$ from the mean value $\bar{h}=0.265$.
    (a) Relation between the ground-state concurrence $C_{\mathrm{gs}}^{(\mathrm{on})}$ and the transition matrix element $\mu_1^{(\mathrm{on})}$.
    The dashed line represents $C_{\mathrm{gs}}^{(\mathrm{on})}=\frac{1-\bar{h}}{1+\bar{h}}\mu_1^{(\mathrm{on})}$ obtained from second-order perturbation theory.
    (b) Relation between the first-excited-state concurrence $C_{\mathrm{1st}}^{(\mathrm{on})}$ and the transition matrix element $\mu_1^{(\mathrm{on})}$.
    The dashed line represents $C_{\mathrm{1st}}^{(\mathrm{on})}=\frac{1+\bar{h}}{1-\bar{h}}\mu_1^{(\mathrm{on})}$.
    }
    \label{fig:random_parameter_concurrence_mu}
\end{figure*}

The above analysis suggests that when $h$ varies over a sufficiently narrow range, the proportionality coefficient between concurrence and the transition matrix element changes only slightly for both the ground state and first excited state. To verify this numerically, $b_x$ and $h$ were independently sampled from uniform distributions over specified ranges, and the Hamiltonian was exactly diagonalized for each parameter set. The resulting relations between concurrence and the transition matrix element are shown in Fig.~\ref{fig:random_parameter_concurrence_mu}.

As shown in Fig.~\ref{fig:random_parameter_concurrence_mu}, for both the ground state and first excited state, the concurrences obtained by exact diagonalization exhibit an approximately linear dependence on the transition matrix element. In addition, the points are distributed near the second-order perturbative predictions evaluated at $\bar{h}=0.265$.

From the above analysis, we find that, in the weak-transverse-field regime, the concurrences of both low-energy eigenstates and the local transition matrix element between them arise at $O(b_x^2)$ from the same transverse-field-induced state mixing. Therefore, along a one-parameter path with fixed $h$, the two quantities are proportional at leading order.

\subsection{Numerical results along the experimental parameter path}
\begin{figure*}[t]
    \centering
    \includegraphics[
        width=0.85\textwidth]{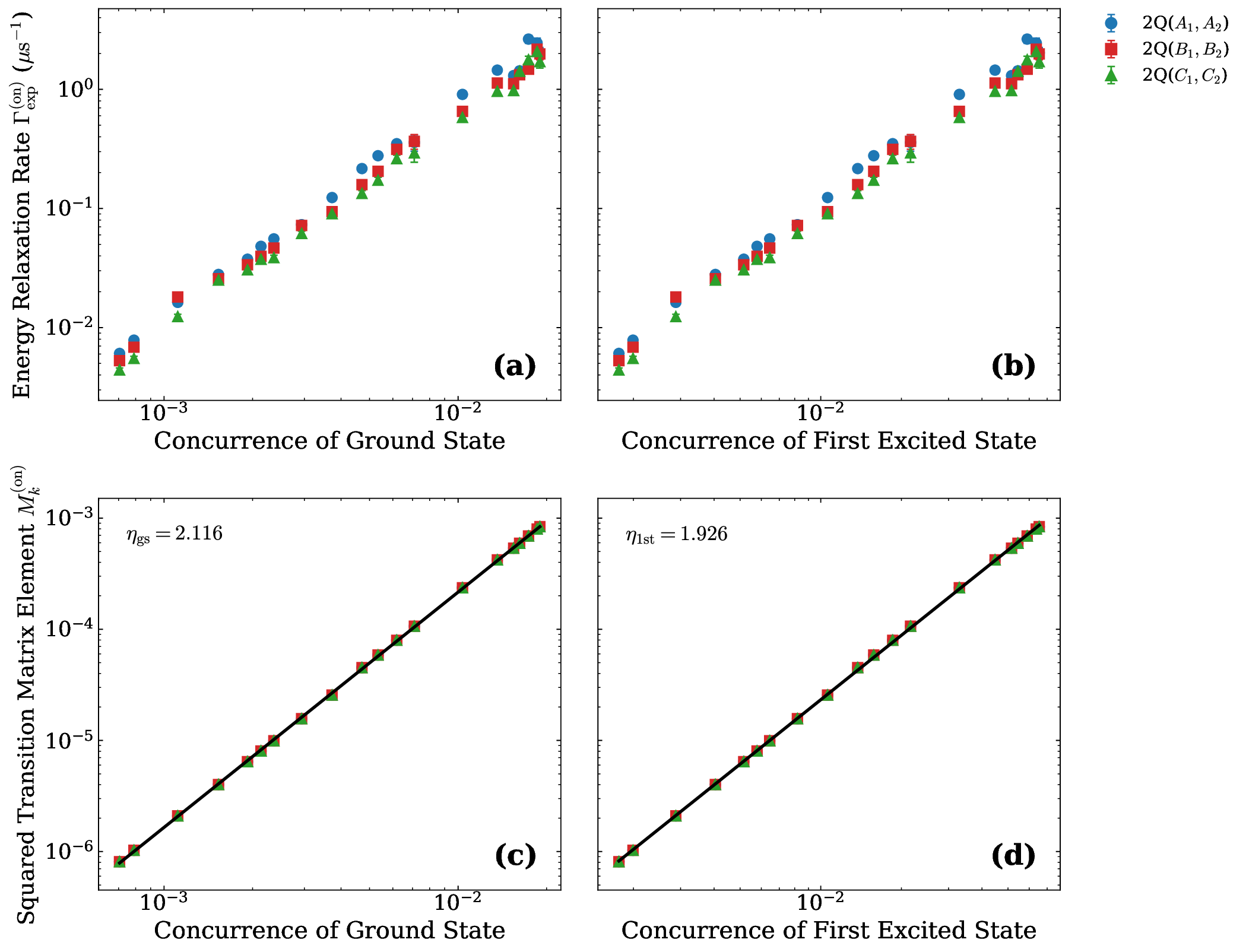}
\caption{
Relationships among the concurrence, experimentally measured energy relaxation rate, and squared transition matrix element in the interacting two-qubit system.
(a) and (b) show the experimentally measured energy relaxation rate $\Gamma^{(\mathrm{on})}_{\mathrm{exp}}$ as a function of the concurrence of the ground state and first excited state, respectively.
(c) and (d) show the squared transition matrix element $M_k^{(\mathrm{on})}$ as a function of the concurrence of the ground state and first excited state, respectively.
The solid lines in (c) and (d) represent the power-law fits obtained using Eq.~\eqref{eq:M_concurrence_fit}.
The fitted exponents are $\eta_{\mathrm{gs}}=2.116$ for the ground state and $\eta_{\mathrm{1st}}=1.926$ for the first excited state.
}
\label{fig:2Q_concurrence_vs_transition}
\end{figure*}

Based on the above analysis, for each interacting measurement condition listed in Table~\ref{tab:measurement_params}(b), we numerically diagonalized the ideal Hamiltonian in Eq.~\eqref{eq:H_on} at the hold point $s=s_d$ and evaluated the concurrences of its ground and first excited states. The corresponding dimensionless transverse-field parameter is $b_x =-A(s_d)/B(s_d)$. The concurrence considered here characterizes the pure eigenstates of the ideal Hamiltonian corresponding to each experimental condition. The relaxation rates shown in Fig.~\ref{fig:2Q_concurrence_vs_transition} are the experimentally measured rates discussed in Sec.~IV.

For all interacting measurement conditions listed in Table~\ref{tab:measurement_params}(b), the concurrences of both the ground state and first excited state of the ideal Hamiltonian were nonzero. Therefore, the corresponding low-energy pure eigenstates of the ideal Hamiltonian are entangled.

As shown in Fig.~\ref{fig:2Q_concurrence_vs_transition}(a) and Fig.~\ref{fig:2Q_concurrence_vs_transition}(b), along the experimental parameter path considered in this study, the experimentally measured relaxation rate tends to increase as the concurrence of the ground state or first excited state increases.
Thus, along the parameter path corresponding to the experimental conditions of this study, the eigenstate concurrence and the experimentally measured energy relaxation rate co-vary.
Fig.~\ref{fig:2Q_concurrence_vs_transition}(c) and Fig.~\ref{fig:2Q_concurrence_vs_transition}(d) show the relation between concurrence and the squared transition matrix element. We fit the numerical results to the following power law,
\begin{equation}
M_k^{(\mathrm{on})}
=
a_cC^{\eta}.
\label{eq:M_concurrence_fit}
\end{equation}
Here, $C$ denotes the concurrence of the corresponding eigenstate; for the ground state,
$C=C_{\mathrm{gs}}^{(\mathrm{on})}$, and for the first excited state,
$C=C_{\mathrm{1st}}^{(\mathrm{on})}$. In addition, $a_c$ is the proportionality coefficient and $\eta$ is the scaling exponent.
The power-law fits yield $\eta_{\mathrm{gs}}=2.116$ for the ground state and $\eta_{\mathrm{1st}}=1.926$ for the first excited state.
Therefore, along the specific parameter path considered in this study, an approximate scaling $M_k^{(\mathrm{on})}\propto C^2$ is observed. This is consistent with the perturbative results $C=O(b_x^2)$ and $\mu_k^{(\mathrm{on})}=O(b_x^2)$ in the weak-transverse-field regime.

\section{Parameters used in the measurements} \label{app:parameters}
For reproducibility, the parameters used in the measurements and the calculated transition matrix elements are summarized in Table~\ref{tab:measurement_params}. In the non-interacting system, all data points were measured with 500 shots. In the interacting two-qubit system, the first six data points, corresponding to the region with relatively large squared transition matrix elements, were measured with 1000 shots, while the remaining data points were measured with 500 shots. The energy difference was adjusted to be approximately $\Delta E = 1.5~\mathrm{GHz}$ for both systems.

\begin{table*}[t]
\centering
\caption{Measurement parameters and calculated squared transition matrix elements used for the measurements in the non-interacting and interacting systems.}

\begin{minipage}[t]{0.45\textwidth}
    \centering
    \resizebox{\linewidth}{!}{
    \begin{tabular}{cccccc}
        \toprule
        No. & $M^{(\mathrm{off})}_k$ & $h$ & $s_d$ & $\Delta E$ (GHz) & Shots \\
        \midrule
        1  & 9.157e-04 & 0.428 & 0.567 & 1.499423 & 500 \\
        2  & 8.155e-04 & 0.425 & 0.570 & 1.500329 & 500 \\
        3  & 4.237e-04 & 0.409 & 0.585 & 1.499480 & 500 \\
        4  & 3.490e-04 & 0.404 & 0.590 & 1.499737 & 500 \\
        5  & 2.331e-04 & 0.395 & 0.599 & 1.499339 & 500 \\
        6  & 1.209e-04 & 0.382 & 0.613 & 1.500408 & 500 \\
        7  & 6.553e-05 & 0.371 & 0.625 & 1.499865 & 500 \\
        8  & 4.702e-05 & 0.365 & 0.632 & 1.500382 & 500 \\
        9  & 3.348e-05 & 0.359 & 0.639 & 1.500291 & 500 \\
        10 & 2.369e-05 & 0.354 & 0.645 & 1.500325 & 500 \\
        11 & 1.099e-05 & 0.342 & 0.660 & 1.500646 & 500 \\
        12 & 8.093e-06 & 0.338 & 0.665 & 1.500150 & 500 \\
        13 & 5.289e-06 & 0.332 & 0.673 & 1.500527 & 500 \\
        14 & 3.299e-06 & 0.326 & 0.681 & 1.500168 & 500 \\
        15 & 1.183e-06 & 0.314 & 0.698 & 1.500489 & 500 \\
        \bottomrule
    \end{tabular}}

    \vspace{2mm}
    \textbf{(a) Non-interacting system}
\end{minipage}
\hfill
\begin{minipage}[t]{0.45\textwidth}
    \centering
    \resizebox{\linewidth}{!}{%
    \begin{tabular}{cccccc}
        \toprule
        No. & $M^{(\mathrm{on})}_k$ & $h$ & $s_d$ & $\Delta E$ (GHz) & Shots \\
        \midrule
        1  & 8.365e-04 & 0.3060 & 0.4420 & 1.499860 & 1000 \\
        2  & 7.976e-04 & 0.3050 & 0.4430 & 1.499856 & 1000 \\
        3  & 6.895e-04 & 0.3020 & 0.4460 & 1.499673 & 1000 \\
        4  & 5.937e-04 & 0.2990 & 0.4490 & 1.499233 & 1000 \\
        5  & 5.361e-04 & 0.2970 & 0.4510 & 1.498796 & 1000 \\
        6  & 4.205e-04 & 0.2930 & 0.4560 & 1.500158 & 1000 \\
        7  & 2.363e-04 & 0.2840 & 0.4660 & 1.500246 & 500 \\
        8  & 1.064e-04 & 0.2730 & 0.4790 & 1.500190 & 500 \\
        9  & 7.965e-05 & 0.2690 & 0.4840 & 1.500282 & 500 \\
        10 & 5.860e-05 & 0.2650 & 0.4890 & 1.499827 & 500 \\
        11 & 4.513e-05 & 0.2620 & 0.4930 & 1.500212 & 500 \\
        12 & 2.559e-05 & 0.2560 & 0.5010 & 1.499997 & 500 \\
        13 & 1.568e-05 & 0.2510 & 0.5080 & 1.500167 & 500 \\
        14 & 9.918e-06 & 0.2460 & 0.5150 & 1.499338 & 500 \\
        15 & 8.032e-06 & 0.2440 & 0.5180 & 1.499572 & 500 \\
        16 & 6.443e-06 & 0.2420 & 0.5210 & 1.499646 & 500 \\
        17 & 4.013e-06 & 0.2380 & 0.5270 & 1.499309 & 500 \\
        18 & 2.096e-06 & 0.2330 & 0.5350 & 1.499974 & 500 \\
        19 & 1.026e-06 & 0.2270 & 0.5450 & 1.500900 & 500 \\
        20 & 8.111e-07 & 0.2250 & 0.5480 & 1.499533 & 500 \\
        \bottomrule
    \end{tabular}%
    }

    \vspace{2mm}
    \textbf{(b) Interacting system}
\end{minipage}
\label{tab:measurement_params}

\end{table*}

\end{document}